\documentclass[12pt,preprint]{aastex}

\newcommand{\etal}{{\it et al.}}

\newcommand{\arcm}{{$^\prime\,$}}
\newcommand{\arcs}{{$^{\prime\prime}\,$}}    
\newcommand{\z}{{z^\prime}}    

\newcommand{\CXO}{{\it Chandra}}

\begin{document}

\title{First Results On Shear-Selected Clusters From the Deep Lens
  Survey: Optical Imaging, Spectroscopy, and X-ray Followup}
 
\shorttitle{Shear-selected Clusters From the DLS}

\author{D. Wittman\altaffilmark{1}, 
I.~P. Dell'Antonio\altaffilmark{2}, 
J.~P. Hughes\altaffilmark{3},
V. E. Margoniner\altaffilmark{1},
J.~A. Tyson\altaffilmark{1}, 
J.~G. Cohen\altaffilmark{4},
D. Norman\altaffilmark{5}}

\altaffiltext{1}{Physics Department, University of California, Davis,
CA 95616; dwittman,vem,tyson@physics.ucdavis.edu}
\altaffiltext{2}{Physics Department, Brown University, Providence, RI
02912; ian@het.brown.edu} 
\altaffiltext{3}{Physics Department, Rutgers University, Piscataway,
NJ 08854; jph@physics.rutgers.edu} \altaffiltext{4}{California
Institute of Technology, Pasadena, CA 91125; jlc@astro.caltech.edu}
\altaffiltext{5}{Cerro Tololo Interamerican Observatory;
dnorman@ctio.noao.edu}

\begin{abstract}

We present the first sample of galaxy clusters selected on the basis
of their weak gravitational lensing shear.  The shear induced by a
cluster is a function of its mass profile and its redshift relative to
the background galaxies being sheared; in contrast to more traditional
methods of selecting clusters, shear selection does not depend on the
cluster's star formation history, baryon content, or dynamical state.
Because mass is the property of clusters which provides constraints on
cosmological parameters, the dependence on these other parameters
could induce potentially important biases in traditionally-selected
samples.  Comparison of a shear-selected sample with optically and
X-ray selected samples is therefore of great importance.  Here we
present the first step toward a new shear-selected sample: the
selection of cluster candidates from the first 8.6 deg$^2$ of the 20
deg$^2$ Deep Lens Survey (DLS), and tabulation of their basic
properties such as redshifts and optical and X-ray counterparts.

\end{abstract} 

\section{Introduction}

The potential of using galaxy clusters as precision cosmological
probes is by now well known (Haiman, Mohr \& Holder 2001).  Clusters
can however be selected by a variety of methods, and the biases of the
different methods are poorly understood.  Traditional methods use
galaxy overdensity or surface brightness enhancements in optical
imaging (Postman \etal\ 1996, Gal \etal\ 2000, Gonzalez \etal\ 2001),
perhaps including color information (Gladders \& Yee 2000 and
subsequent work); or X-ray emission (Rosati, Borgani \& Norman 2002).  Newer
methods which have yet to produce sizable samples include the
Sunyaev-Zel'dovich effect (SZE; Carlstrom, Holder \& Reese 2002) and
weak gravitational lensing (Tyson, Valdes \& Wenk 1990; Schneider
1996; Wittman \etal\ 2001, 2003).

The types of biases which could exist are clear, even if their
practical impact is not.  Most methods, except X-ray, involve an
integral of a density along the line of sight, making them susceptible
to line-of-sight projections.  However, the impact of false positives
from projections can be greatly mitigated by spectroscopic followup,
which is often required for a redshift in any case.  X-ray selection
is less sensitive to projections because the emission is proportional
to the square of the local density, but that could also lead to biases
due to internal substructure or asphericity.  Most methods, except
lensing, involve trace constituents of the cluster whose connection to
the underlying predictable (and cosmologically significant) quantity,
mass, is not completely understood.  Optical, X-ray, and SZE selection
depend on baryon content, which is a minor cluster component compared
to dark matter.  This alone is probably not a serious bias, because in
a bottom-up structure formation scenario, large dark matter
concentrations without baryons are very unlikely.  However, optical
selection additionally depends on star formation history, and X-ray
selection depends on the heating of the intracluster medium (ICM).
Weinberg \& Kamionkowski (2002) predict that up to 20\% of
weak-lensing clusters have not yet heated their ICM to a level
detectable with current X-ray missions.  They predict this fraction to
be independent of the signal-to-noise of the lensing detection, but
increasing with redshift.

The redshift dependences also vary greatly across the methods.
Optical and X-ray selection depend on the luminosity distance of the
cluster and on k-corrections.  SZE has the special property of being
redshift-independent. This allows it to reach to very high redshift,
where the cluster abundance is very sensitive to the cosmology, but it
may increase the opportunities for chance projections.  Lensing
occupies a middle ground, with a broad sensitivity to mass at
redshifts $\sim$ 0.2---0.7 for typical surveys.  Clearly, no one
method by itself will provide a perfect sample, and we must work to
understand the biases through extensive intercomparisons.  Some work
has been done on optical-X-ray comparison (Donahue \etal\ 2001, 2002), 
but the advent of shear and SZE selected samples introduces a new
challenge and a new opportunity to deepen our understanding.

Here we present the first shear-selected cluster sample, selected from
8.6 deg$^2$ of the Deep Lens Survey (DLS; Wittman \etal\ 2002).  The
DLS is a deep ground-based $BVR\z$ imaging survey of 20 deg$^2$ with
consistent good ($<0.9$\arcs) image quality in $R$, where the source
galaxy shapes are measured.  The rest of the paper is organized as
follows: in Section~\ref{sec-hist}, we give a brief history of shear
selection; in Section~\ref{sec-select}, we describe our data and
methods; in Section~\ref{sec-cand}, we present the cluster candidates
individually; and in Section~\ref{sec-summary}, we offer a summary and
discussion.  Throughout, we use ``cluster'' to mean any mass
concentration, without implying any properties of member galaxies.

\section{History of Shear Selection}
\label{sec-hist}

Tyson, Valdes \& Wenk (1990) first reported the systematic alignments
of background galaxies around a foreground cluster, now referred to as
weak lensing.  At the time, CCD fields of view were small and it was
infeasible to search for new clusters with this method, limiting it to
followup of already-known clusters.  Tyson (1992) was
the first to suggest using weak lensing to search for mass
concentrations, rather than simply following up known clusters.  Early
work on cosmology constraints from cluster counts assumed NFW profiles
for all clusters (Kruse \& Schneider 1999).  Bartelmann, King \&
Schneider (2001) then pointed out that because the profile has a big
impact on detectability, cluster counts may reveal as much about
dark-matter profiles as about global cosmological parameters.

However, in a larger view of things, cluster profiles are directly
related to the cosmology and dark matter model.  Shear-selected
cluster counts are still straightforwardly derivable given the
cosmological model, even if one has to perform an n-body simulation to
realistically model the effects of cluster profiles.  The first work
in this direction was that of White, van Waerbeke \& Mackey (2001),
followed by Hennawi \& Spergel (2004), who performed
ray-tracing through a large-scale particle mesh simulation and
computed the resulting ``observed'' mass and redshift distribution of
clusters found in a mock survey.  Both groups found that
shear-selected samples will always have false positives, even for very
high shear thresholds.  These are caused by large-scale structure
noise, which cannot be overcome by improved observations.  
Neither group addressed the feasibility of deriving constraints on
cosmological parameters by including ``false positives'' in both
observations and simulations; that is, simply measuring the abundance
of shear peaks, which is easily computed from n-body simulations.
Such an approach would preserve the clean comparison with theory which
is the virtue of shear selection, but on the other hand, much of a
cluster survey's power to distinguish cosmologies comes from the
redshift distribution.  How to balance these factors to maximize the
information from shear-selected clusters remains an open issue.

Meanwhile, the first claims of clusters discovered via weak lensing
were published.  In many of them (Erben \etal\ 2000; Umetsu \&
Futamase 2000; Clowe, Trentham \& Tonry 2001; Miralles \etal\ 2002),
the interpretation of the observations is not clear because the object
causing the shear has not been identified with a redshift, without
which mass and mass-to-light ratios (M/L) cannot be computed.  The
first shear-selected mass with a spectroscopic redshift appeared in
2001 ($z=0.27$, Wittman \etal\ 2001), followed in 2003 by another, an
early result from the DLS ($z=0.68$, Wittman \etal\ 2003).  The same
year, Dahle \etal\ (2003) and Schirmer \etal\ (2003) each identified
several shear-selected masses with redshifts roughly determined from
two-color photometry ($z \sim 0.5$).  Most of these clusters were
serendipitous, usually near X-ray selected clusters which were the
main target of the observations.  A truly representative sample can
only be taken from a survey of an unbiased area.  The first published
survey results were those of Miyazaki \etal\ (2002), who counted
convergence peaks in a 2.1 deg$^2$ Subaru field, but attempted no
followup in terms of redshifts, member galaxies, or X-ray emission.

Wittman \etal\ (2001) introduced the idea of tomographic confirmation.
That is, the shear around a given lens must grow as a specific
function of source redshift.  To properly interpret a peak on a
convergence map, one must confirm that the redshift dependence is as
expected.  Otherwise, the ``shear'' could be due to systematics in the
data, such as optical distortion or local variations in the
point-spread function.  One might expect that this would also serve as
a check against false positives from projections, but Hennawi \&
Spergel (2004) found false shear-selected candidates in their
simulations which displayed perfectly sensible shear-redshift curves.
It is understandable, given the broadness of the lensing redshift
dependence and the imprecision of photometric redshifts, that
filaments seen end-on cannot be diagnosed from the lensing information
alone.  Although it is not clear how often false positives remain with
tomography, it is natural to examine other followup possibilities to
weed out the false positives.

The use of spectroscopic or X-ray confirmation would reintroduce some
of the biases of those methods, but perhaps at a much reduced level.
For example, spectroscopic confirmation may be possible even for
clusters with high M/L which would have escaped detection in an
optical search.  This approach would still fail on the extreme
scenario of a pure dark matter cluster, but it seems likely that there
is a continuum of M/L, and this approach at least allows us to go much
further down the continuum than before.  X-ray confirmation may not be
foolproof, given the prediction that a significant fraction of weak
lenses are X-ray dark, regardless of the signal-to-noise of the
lensing detection.  In fact, Weinberg \& Kamionkowski (2003) argue for
using the X-ray dark fraction in lensing samples as a constraint on
dark energy which, unlike raw counts, is robust against uncertainties
in observational thresholds.

These are open issues.  In this paper, we try to shed light on the
matter by providing critical new data: a shear-selected sample from
real observations.

\section{Cluster Selection Procedure}
\label{sec-select}

\subsection{Optical Imaging Data}

The DLS consists of five well-separated 2$^\circ \times 2^\circ$
fields (see Table~\ref{table-fieldcoords} for coordinates; all
coordinates in this paper are J2000).  The northern fields (F1 and F2)
were observed using the Kitt Peak Mayall 4-m telescope and Mosaic
prime-focus imager (Muller \etal\ 1998).  The southern fields (F3
through F5) were observed with a similar setup at the Cerro Tololo
Blanco 4-m telescope.  The Mosaic imagers consist of a 4$\times$2
array of three-edge-buttable 2k$\times$4k CCDs, providing a 35\arcm\
field of view with 0.26\arcs\ pixels and minimal gaps between the
devices.  Each DLS field is divided into a 3$\times$3 grid of
40\arcm$\times$ 40\arcm\ subfields.  These subfields are slightly
larger than the Mosaic field of view, but are synthesized with
dithers of up to 800 pixels (208\arcs).  The primary motivation for
the large dithers is to provide good sky flats.

\begin{table}
\begin{center}
\begin{tabular}{|c|c|c|c|c|}
\hline
Field & RA\tablenotemark{a} & DEC & l,b & E(B-V)\tablenotemark{b} \\ \hline
F1 & 00:53:25.3  &    +12:33:55   &    125,-50 & 0.06 \\
F2 & 09:18:00    &    +30:00:00   &    197, 44 & 0.02 \\
F3 & 05:20:00    &    -49:00:00   &    255,-35 & 0.02 \\
F4 & 10:52:00    &    -05:00:00   &    257,47  & 0.025 \\
F5 & 13:55:00    &    -10:00:00   &    328,49  & 0.05 \\ \hline
\end{tabular}
\end{center}
\caption{DLS field information.}
\tablenotetext{a}{J2000, field center.}
\tablenotetext{b}{From Schlegel, Finkbeiner \& Davis (1998).  The
  value given is an average over each 4 deg$^2$ field.}
\label{table-fieldcoords}
\end{table}

The planned final depth for each subfield is twenty 600-s exposures in
$B$, $V$, and $\z$, and twenty 900-s exposures in $R$.  The dithers
are large enough to overlap adjacent subfields, providing more uniform
depth at the subfield edges and good astrometric and photometric
tie-ins which allow construction of a uniform catalog covering the
entire 2$^\circ \times 2^\circ$ field.

A key observing strategy of the DLS is to observe in $R$ when the
seeing FWHM is $<0.9$\arcs, and in $BV\z$ otherwise.  Thus, at the end
of the survey, the $R$ band imaging will have fairly uniform good
resolution, as well as greater depth due to longer exposure time and
greater system sensitivity in $R$.  Shape measurements are thus done
only in $R$ band, with $BVz$ used to provide color information for
photometric redshifts.  See Wittman \etal\ (2002) for further details
regarding the survey design, field selection, etc.  Note that
photometric redshifts are not generally used in this paper because
they were not available at the time the clusters were selected for
X-ray followup, and they are not precise enough to rule out some types
of projections.  However, they will be used in future papers exploring
tomography of these clusters, and for measuring the source redshift
distribution for mass calibration.

Observing began in November 1999 at Kitt Peak, and in March 2000 at
Cerro Tololo.  The imaging data used for cluster selection in this
paper cover 10.7 deg$^2$ in fields F2 through F5, which had reached a
cumulative exposure time of at least 9000 s in $R$ as of March 2002,
when the cluster sample was selected for X-ray followup.  The
effective area searched is somewhat less due to exclusion of edge
areas (see below).  $BV\z$ data were not used in the selection
due to significant gaps in coverage at the time of selection.
However, $BV\z$ data are now available, and we are able to present
true-color images of each candidate below.

\subsection{Image Processing}

We remove instrumental artifacts such as bias, flatfield, etc., and
perform astrometric calibration, in a standard way with the IRAF
package {\it mscred}.  See Wittman \etal\ (2002) for further technical
details regarding these steps.  We then make a stacked image of each
subfield in $R$ as follows.

\begin{enumerate}

\item For each device in each contributing exposure, make a quick
  catalog of the high signal-to-noise objects.  The initial step uses
  SExtractor (Bertin \& Arnouts 1996), after which we cut on semiminor
  axis to eliminate cosmic rays; eliminate saturated objects; convert
  pixel positions to equatorial coordinates using routines from the
  wcstools library (Mink 2000), as SExtractor does not read the TNX
  coordinate system used by Mosaic; and compute the adaptive moments
  using the {\tt ellipto} program (Bernstein \& Jarvis 2002). The
  adaptive moments are second central moments weighted by a matched
  elliptical Gaussian and are equivalent to finding the best-fit
  elliptical Gaussian for each object.  This is not used for shear
  measurement at this stage, but merely to aid in identification of
  stars, whose adaptive moments are not magnitude dependent as are the
  SExtractor intensity-weighted moments computed within a limiting
  isophote.  The initial SExtractor step also produces a
  sky-subtracted image which will be the real input to the stacking
  software, {\it dlscombine}.  This removes any need to match the sky
  levels when stacking, and also prevents the sky from becoming
  nonuniform during the non-flux-conserving repixelization step
  (below).

{\item The {\it mscred} astrometric calibration is not good enough to
  stack images directly; we find shifts of typically $\sim$ 0.04\arcs\
  between the astrometry of overlapping exposures, which would lead to
  spurious stretching of galaxy shapes.  Therefore we match all the
  catalogs in equatorial coordinates to produce a master catalog which
  defines the astrometry of the final stack image.  All subsequent
  coordinate transformations are derived by matching individual
  exposures against this master, which reduces the offsets to zero
  within an uncertainty of $\sim$0.004\arcs.  The master position of
  an object is simply the mean RA and DEC at which it was observed;
  objects observed only once (within the tolerance of 1.8\arcs) are
  dropped as being possibly spurious. The mean magnitude is also
  recorded.\label{item-mastercat}}

\item Transform the master catalog positions to desired pixel
  coordinates in the output stack image.  The coordinate system of the
  stack is a simple tangent plane projection with no optical
  distortion.

\item For each device in each exposure, determine the transformation
  from input pixel coordinates to output pixel coordinates.  We use a
  third-order polynomial.  Together, this and the previous two steps
  insure the best possible image registration, robust against small
  errors in the astrometric calibration of each input image.  Precise
  registration is important, as errors could mimic spurious shear.
  Typical rms residuals of an input image when matched against the
  USNO A-1 catalog (Monet \etal\ 1998) are $\sim$ 0.35\arcs, whereas
  the residuals matched against the master catalog or the stacked
  image are $\sim$0.03\arcs.

\item For each device in each exposure, identify stars based on their
  position in the magnitude-size diagram, where size is defined in
  Bernstein \& Jarvis (2002) as the sum of the adaptive moments
  $I_{xx} + I_{yy}$ times a correction factor for non-Gaussianity.
  Initial identification is done with an automatic algorithm which
  identifies the typical stellar size by looking for a peak in the
  size histogram, then selects the magnitude range for which there is
  a significant density enhancement at that size, compared to a
  control region at larger size.  We inspect all selections, and
  manually adjust the selection in $\sim$5\% of the exposures,
  typically because a few unflagged saturated stars caused the
  algorithm to identify a wider than necessary stellar locus.  A
  typical diagram is shown in Figure~\ref{fig-starselect}, with the
  eight devices shown separately.

\clearpage\begin{figure}
\centerline{\resizebox{3in}{!}{\includegraphics{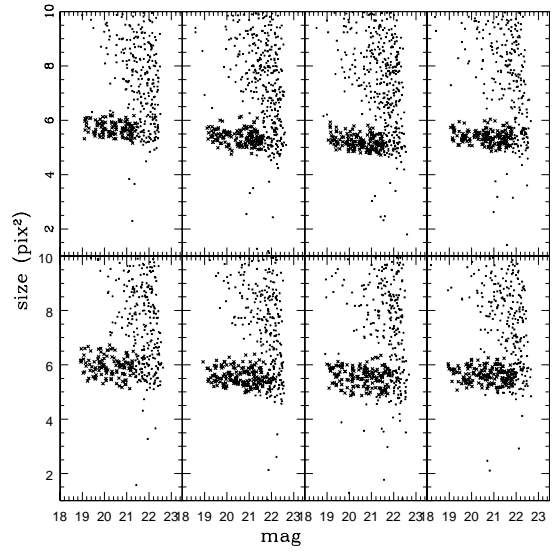}}}
\caption{Typical star selection scenario in $R$ band. All objects in
  the image are shown with lighter points, and the selected stars are
  shown in heavier points.\label{fig-starselect}}
\end{figure}\clearpage

\item For each device in each exposure, analytically ``undistort'' the
  PSF adaptive moment combinations $I_{xx} - I_{yy}$, $I_{xy}$, and
  $I_{xx} + I_{yy}$ using the above coordinate transformation, then
  fit a smooth function to their spatial variation.  This is
  mathematically equivalent to, but computationally faster than,
  repixelizing the image to remove the optical distortion and then
  measuring the PSF shape in the distortion-free coordinate system.
  The fitting function used is a third-order polynomial in x and y
  pixel coordinates.  ``Stars'' which lie outside any of these fits by
  3$\sigma$ or more are rejected as likely interloping galaxies.
  These fits are used to derive a PSF circularization kernel as in
  Fischer \& Tyson (1997), although it is not applied immediately.

\item For each device in each exposure, correct the photometry in its
  catalog for the following effect.  Pixels at the corner of the
  Mosaic subtend $\sim$5\% more area than pixels at the center, due to
  optical distortion.  Therefore, these pixels collect more sky
  photons, and objects which fall on these pixels are unfairly
  penalized when sky flats are applied, even though the sky appears to
  be flat after application.  These objects will regain their lost
  flux when {\it dlscombine} remaps the image to the distortion free
  coordinate system {\it without} flux conservation.  However, to
  correctly determine photometric offsets from the catalogs, the
  correction also has to be applied to any catalogs made before the
  remapping step.

\item Make a master photometric catalog from all the corrected
  catalogs, following the same rules as in step~\ref{item-mastercat}.

\item For each device in each exposure, derive a relative photometric
  offset by matching to the master catalog and computing the
  3$\sigma$-clipped mean of the magnitude differences of the matching
  objects.  In more recent versions of the pipeline, this and the
  previous step have been combined into an iterative search for the
  set of offsets which produce the most uniform master catalog.  Also,
  all devices in a given exposure are now handled together, which
  provides extra robustness against wrong solutions for devices which
  have lost a large fraction of area to a bright star or readout
  problem.

\end{enumerate}

The PSF-related steps are skipped for $BV\z$, but otherwise the
stacking procedure is the same.  We do not make a single 2$^\circ
\times 2^\circ$ image because of its prohibitive file size and
increased distortion due to tangent plane projection.  Rather, we make
10k$\times$10k (43\arcm$\times$43\arcm) stacked images of each
subfield/filter combination, providing 6\arcm\ of overlap between
adjacent subfields, and combine the subfield catalogs afterward.  For
stacks with this footprint, many contributing exposures come from
adjacent subfields.  Thus, the overlap regions of the stacks do not
represent independent data, but do represent the effects of differing
pixelizations (each stack is a tangent projection about its center)
and different large-scale PSF fits.

We then run {\it dlscombine}, which iterates over pixels in
the output image, applying bad pixel masks, the PSF circularization
kernel, coordinate transformations (with sinc interpolation),
and photometric offsets to each relevant input image, and computes the
mean of the valid contributing pixels, with a 3$\sigma$ clipping.
Examination of the sky noise as a function of the number of input
images reveals $\sqrt{n}$ improvement, indicating that we are
successfully removing instrumental artifacts and reaching the Poisson
noise limit.

\begin{figure}
\centerline{\resizebox{3in}{!}{\includegraphics{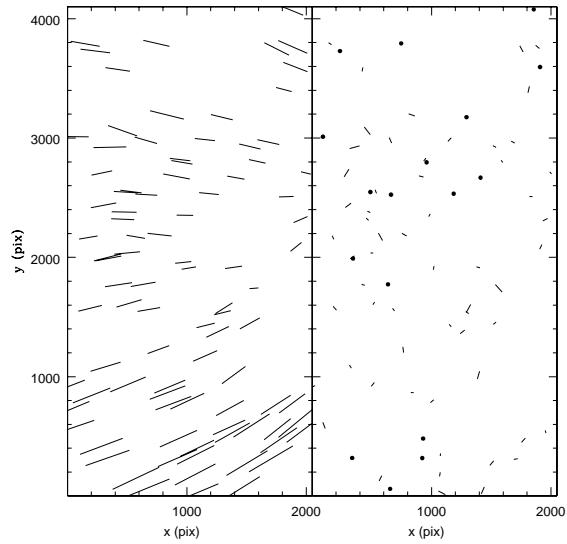}}}
\caption{PSF circularization of one device from one exposure
(9\arcm$\times$18\arcm).  Typical ellipticity amplitudes before
circularization are up to 6\% with strong spatial correlations. After
circularization, this falls to 1\% or less, with very weak spatial
correlations, and improves further upon combining multiple exposures.
\label{fig-circ}}
\end{figure}

An example of the performance of the Fischer \& Tyson circularization
kernel is shown in Figure~\ref{fig-circ}.  Typical ellipticity
amplitudes before circularization are up to 6\% with strong spatial
correlations, falling to 1\% or less, with very weak spatial
correlations, after application of the kernel.  Stacking multiple
exposures further reduces correlations, as exposures separated by as
little as fifteen minutes in time often have different PSF patterns.
After stacking, the mean PSF ellipticity, averaged over the 43\arcm\
$\times$ 43\arcm\ area of the stack, is typically $\sim$0.1\% in each
component.  We then apply another round of circularization to the
stack, to smooth out any errors introduced by stacking errors, or by
the inability of the relatively small (3$\times$3 pixel) kernel to
handle large and/or highly elliptical PSFs in one pass.  After this
stage, the typical mean PSF ellipticity drops to $\sim$0.01\% in each
component, consistent with zero given the scatter among PSF stars.

\clearpage\clearpage

Stacks were made for all of fields F2 and F4, and contiguous portions
of fields F3 (two subfields) and F5 (four subfields).  The mean image
quality, after all circularizations, was 0.90\arcs\ with an rms
subfield-to-subfield variation of 0.03\arcs.  The total area covered
by the imaging is 10.7 deg$^2$, before excluding edge areas as
discussed below.

\subsection{Convergence Maps}

To make continuous convergence maps larger than the subfield size, we
cataloged each subfield's deep $R$ image separately using SExtractor,
and stitched the subfield catalogs together into one supercatalog for
each field, as follows.  First, exclusion zones along the outer, noisy
edges of the stacks were defined based on manual inspection of the
image and catalog together.  Subfield edges which abutted another
subfield were not irregular; only the outer edges of the field
required exclusion zones.  Then, objects which appeared in only one
subfield catalog (based on RA and DEC) were passed directly on to the
final catalog. Objects in overlap regions were matched in RA and DEC
and their properties subjected to consistency tests.  Objects were
rejected if the multiple measurements disagreed by 0.2$^m$ or more in
aperture magnitude, or by 1.0 pixel$^2$ or more in any of the three
second central moments $I_{xx}, I_{yy}$ or $I_{xy}$.  Output
statistics of the matching program were monitored to insure that
relatively few objects were rejected, and that in a given overlap
region, there were very few orphan objects in one subfield's catalog
but not in the others.  These conditions were satisfied for all
overlaps, indicating that the supercatalogs are free of gaps,
duplications, and discontinuities.

At this point the supercatalogs contained $\sim$250,000 sources per
square degree, with counts peaking at $R=25.5$.  
Adding the requirement that the adaptive moments be successfully
measured by {\tt ellipto} decreased the usable source density by
$\sim$20\%.

The supercatalogs were then filtered to remove low-redshift galaxies
as well as possible given the limited information available at the
time.  We determined magnitude and size cuts by maximizing the
detection signal-to-noise of simulated clusters as well as of the
already-confirmed clusters in the real data (candidates 1 and 8 in
Table~\ref{table-candidates}).

The final cuts were:

\begin{itemize}

\item $23 < R < 25$. This is a typical cut used in lensing,
  representing a balance between removing obvious low-redshift
  galaxies, removing the faintest, noisiest galaxies, and retaining a
  large sample.

\item {\it $5.8 < size < 20$ pix$^2$}, where size is defined above.
With a PSF size of $\sim$6, this included a range of galaxies from
barely resolved or even unresolved, to quite well resolved, but
excluded very large foreground galaxies.  We made no attempt to
calibrate the shear by correcting the observed ellipticities to their
pre-seeing values, as we were most interested in making a ranked list
rather than imposing a physical threshold, and the seeing was
considered uniform enough to fairly rank the shear peaks in the
different fields against each other.  This does have the effect of
downweighting high-redshift clusters, whose background galaxy shears
are more diluted by seeing.  A retrospective seeing correction shows
that this is a small effect, and has no consequences in any case
because in this paper we do not compare the cluster redshift
distribution with expectations from n-body simulations.  The final DLS
shear-selected cluster sample will be based on a shear threshold which
has been corrected for the small field-to-field differences.  Another
potential problem with including some unresolved sources would be
stellar contamination, but the ratio of galaxies to stars at these
magnitude ranges and galactic latitudes is so large that in practice
stars are not a significant factor.

\item {\it isophotal area $ <150$ pixels}.  The intent of the
isophotal area cut is to exclude large foreground galaxies.  Although
degenerate with magnitude and size cuts, the isophotal area cut had
some effect in the simulations, and we therefore applied it to the
data as well.  In practice, this constraint removed about 10\% of the
galaxies which had survived the other cuts, mostly on the bright end
where the limiting isophote of a galaxy could be at a large distance
from its center even if the moments $I_{xx} + I_{yy}$ were small.  For
example, more than half the galaxies with $23 < R < 23.2$ were
rejected by this criterion, compared to 0.5\% of galaxies in the range
$24.8 < R < 25$.  

\end{itemize}

\begin{figure}
\centerline{\resizebox{4in}{!}{\includegraphics{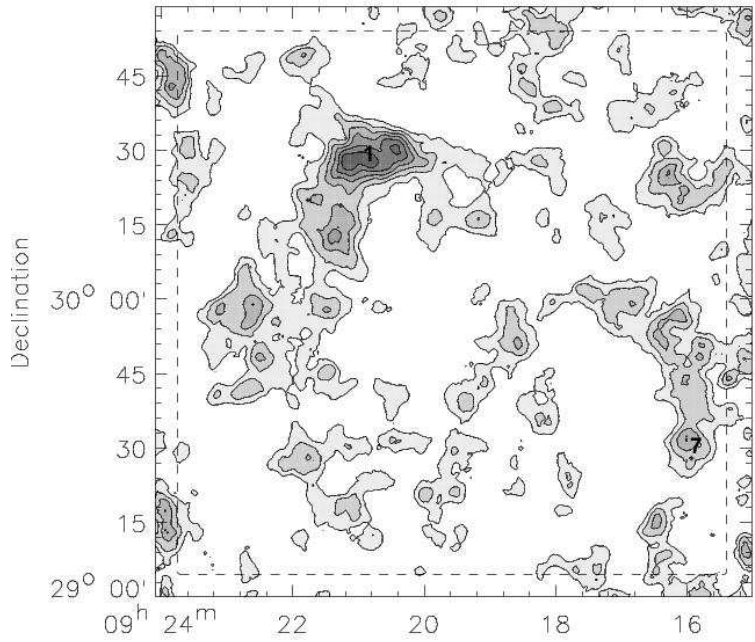}}}
\caption{Convergence map for field F2, covering a full 2$^\circ \times
2^\circ$ area.  For all convergence maps in this paper, white
indicates low surface mass density, black indicates highest density;
north is up, east is left; dashed lines indicate the 5\arcm\ edge
exclusion zone; and labels indicate DLS cluster number.
\label{fig-f2mass}}
\end{figure}

\begin{figure}
\centerline{\resizebox{4in}{!}{\includegraphics{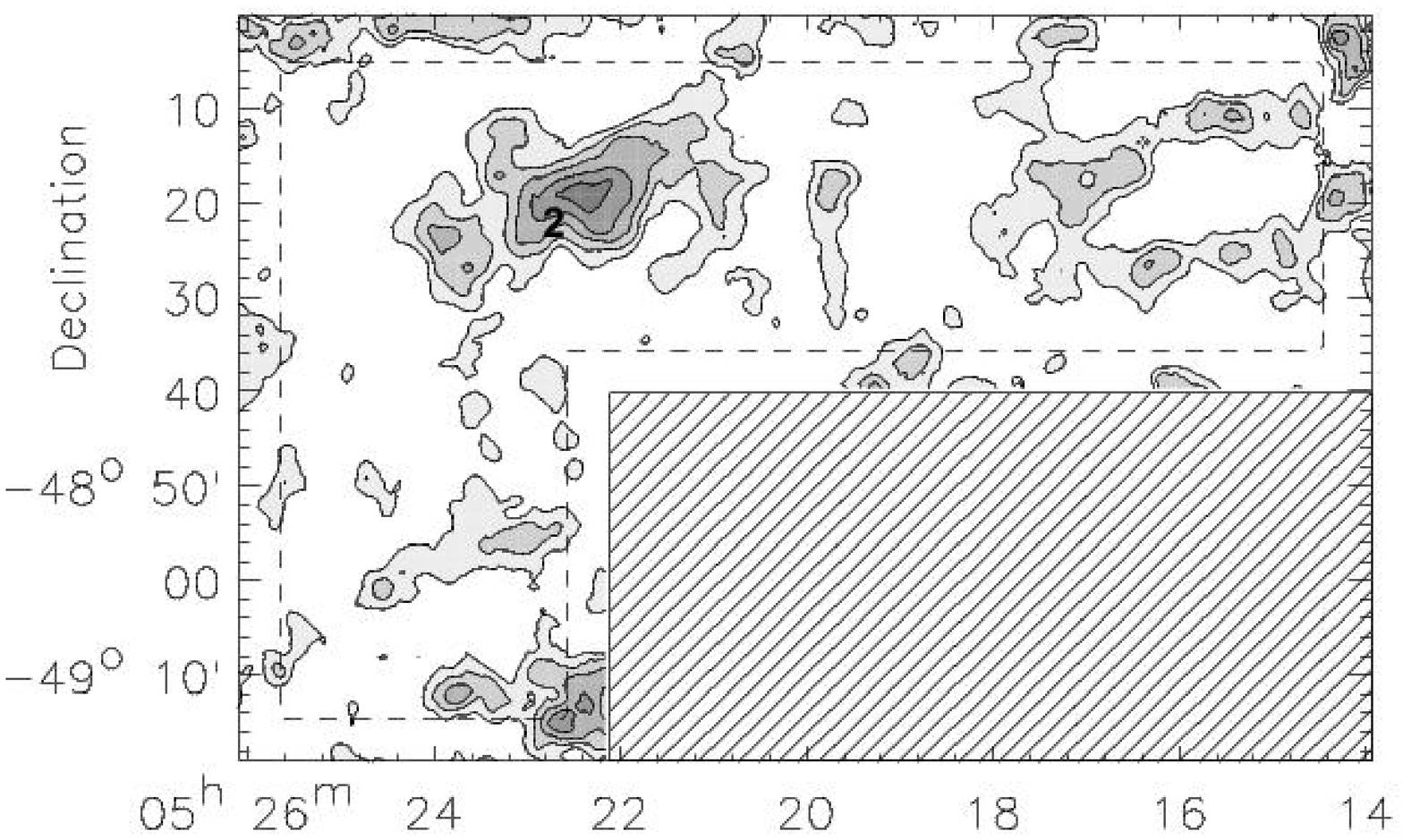}}}
\caption{As for Figure~\ref{fig-f2mass}, but for the portion of field
F3 covered by the selection (1.8 deg$^2$ before edge exclusion).
\label{fig-f3mass}}
\end{figure}

\begin{figure}

\centerline{\resizebox{4in}{!}{\includegraphics{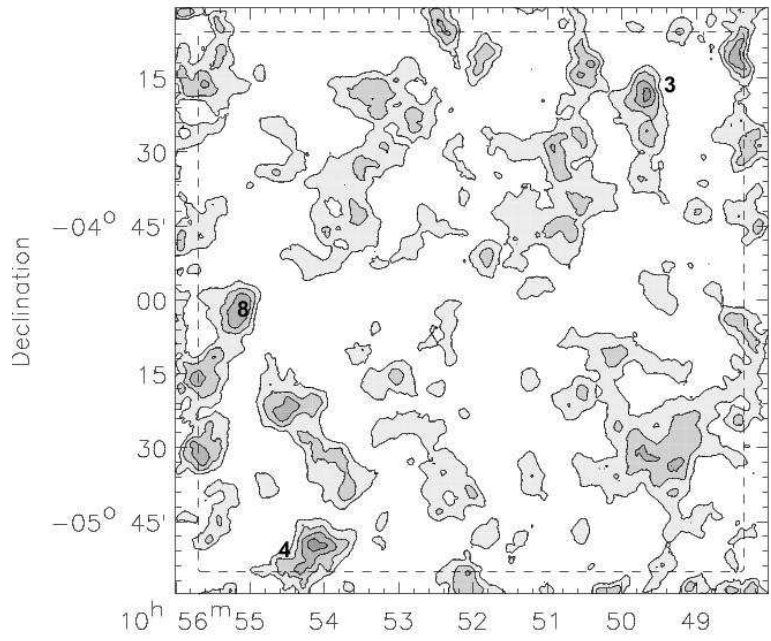}}}
\caption{As for Figure~\ref{fig-f2mass}, but for field F4 (4 deg$^2$
before edge exclusion).
\label{fig-f4mass}}
\end{figure}

\begin{figure}
\centerline{\resizebox{2.667in}{!}{\includegraphics{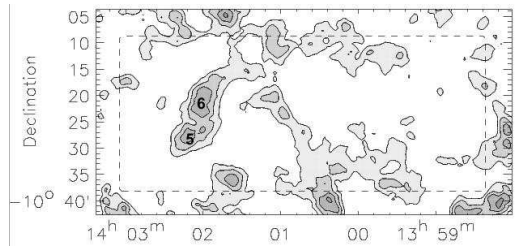}}}
\caption{As for Figure~\ref{fig-f2mass}, but for the portion of field
F5 covered by the selection (0.9 deg$^2$ before edge
exclusion).\label{fig-f5mass}}
\end{figure}

This filtering reduced the size of the catalogs by about 70\%, to
roughly 70,000 sources deg$^{-2}$.  As a control, we cross-correlated
the ellipticities of these sources with those of the stars.  The
result was consistent with zero at all angular scales.  We then
convolved the filtered catalogs with a kernel
of the form
$$ r^{-2} (1-\exp({-r^2 \over 2r_i^2})) \exp({-r^2 \over 2r_o^2})
\eqno{(1)}$$ where $r_i$ and $r_o$ are inner and outer cutoffs,
respectively, to produce unnormalized convergence maps.  This kernel
is a modified version of the kernel presented in Fischer \& Tyson
(1997), with a Gaussian outer cutoff added to suppress noise from
sources at large projected radius, where other structures are adding
noise to the tangential shear field.  We used $r_i = $4.25\arcm\ and
$r_o =$50\arcm.  The results were pixelized onto maps with 30\arcs\
pixels.  These maps are shown in Figures~\ref{fig-f2mass} through
\ref{fig-f5mass}.

\clearpage

\clearpage

\clearpage

\clearpage\clearpage

\subsection{Candidate Identification}

We compiled a ranked list of peaks in the convergence maps and then
eliminated those within 5' of an edge, where the convergence map noise
increases due to lack of input data over much of the filter footprint.
The effective area searched thus decreased to 8.6 deg$^2$.  This
subset of the DLS area has a higher ratio of perimeter to area than
does the full survey.  The same edge cut applied to the full survey
would yield an effective area of 16.8 deg$^2$.

We also made maps where the prefactor in Equation 1 was $r^{-1}$
rather than $r^{-2}$, though with smaller cutoff radii (1.25\arcm\ and
34\arcm) so that the effective amount of smoothing was similar.  These
are akin to gravitational pseudo-potential maps rather than
convergence maps.  Although the two sets of maps used the same input
data, one might expect shot noise and some systematic errors to
propagate somewhat differently through the two algorithms.  We found
that for the top several candidates, the rankings yielded by the two
types of maps were identical.  Below that, the rankings tended to
disagree by a place or two, as there were several candidates with
nearly identical peak values, whose rankings were easily shuffled by a
small change in the noise properties.  In other words, peaks in the
convergence map corresponded to peaks in the pseudo-potential map, and
vice versa.  We conclude that the candidates were robustly detected
(see below for a quantitative estimate).  The pseudo-potential maps
are not presented here because they appear quite similar to the
convergence maps.  The final rankings were determined by the sum of
the convergence and potential maps (normalized to make roughly equal
contributions to the sum), as listed in Table~\ref{table-candidates}.

\begin{figure}
\centerline{\resizebox{2in}{!}{\includegraphics{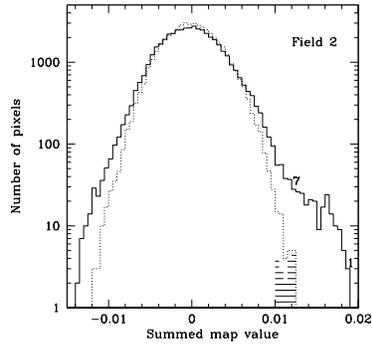}}}
\caption{Histogram of values in the sum map for one field (F2), for
the tangential component of shear (solid outline) and for the other
component as a control (dash outline).  The two candidates in this
field are labeled at their sum map values, above the solid histogram.
The shaded area of the control histogram represents an edge feature
which would have been excluded with a slightly larger edge cut, with
no effect on the actual sample.
\label{fig-pixhist}}
\end{figure}

As a control test, we repeated the entire map-making procedure using
the non-tangential component of the shear.  Figure~\ref{fig-pixhist}
shows the histogram of values in the sum map for one field (F2), for
the tangential component (solid outline) and the other component (dash
outline).  As expected, the distribution is wider in the first case,
reflecting the presence of real clusters and voids.  The values at the
locations of the two cluster candidates in the field, candidates 1 and
7, are labeled above the solid line.  The shaded area of the control
histogram corresponds to a single feature right at the edge of the
trimmed field. In other words, a slightly larger edge exclusion would
have been better.  None of our actual candidates are on the edge, so
this had no effect on the sample.  In summary, our candidates are
above the maximum values in the control maps.

\clearpage\clearpage

The IAU-approved notation for DLS clusters is {\it DLSCL
JHHMM.m+DDMM}, where {\it H,D}, and {\it M} refer to hours, degrees,
and minutes respectively, and {\it m} refers to tenths of minutes of
time.  {\it DLS JHHMMSS.ss+DDMMSS.s}, where {\it S} and {\it s} refer
to seconds and tenths of seconds respectively, is reserved for
individual sources detected in the DLS, which have much higher
positional accuracy.  The sample was defined by a cutoff in the
ranking rather than a cutoff in shear or signal-to-noise because the
purpose here is to define a sample small enough to allow comprehensive
followup with optical spectroscopy and X-ray spectro-imaging.  The
cutoff in Table~\ref{table-candidates} reflects the cutoff in actual
X-ray followup.  The final DLS shear-selected cluster sample may go
further down the rankings, or equivalently to a lower shear level.  As
a quantitative estimate of the cutoff signal-to-noise ratio (SNR), we
note that one of the lowest-ranked candidates presented below has
already been published by Wittman \etal\ (2003) as a 3.7$\sigma$
detection.

\begin{table}
\begin{center}
\begin{tabular}{|l|l|l|l|l|l|}
\hline
Rank & ID & RA(J2000) & Dec(J2000) & Peak value\tablenotemark{a} & Field\\
\hline
1 & DLSCL J0920.1+3029 & 09:20:08 & +30:29:53 & 0.0188  & 2 \\
2 & DLSCL J0522.2-4820 & 05:22:17 & -48:20:10 &  0.0151 & 3 \\
3 & DLSCL J1049.6-0417 & 10:49:41 & -04:17:44  & 0.0136 & 4\\
4 & DLSCL J1054.1-0549 & 10:54:08 & -05:49:44  & 0.0125 & 4\\
5 & DLSCL J1402.2-1028 & 14:02:12 & -10:28:14  & 0.0123 & 5\\
6 & DLSCL J1402.0-1019 & 14:02:03 & -10:19:44  & 0.0120 & 5\\
7 & DLSCL J0916.0+2931 & 09:16:00 & +29:31:34  & 0.0119 & 2\\
8 & DLSCL J1055.2-0503 & 10:55:12 & -05:03:43  & 0.0119 & 4\\
\hline
\end{tabular}
\end{center}
\caption{Ranked list of cluster candidates.\label{table-candidates}}
\tablenotemark{a}{Sum of convergence map and pseudo-potential map in
  arbitrary units.}
\end{table}

Note that the clusters cannot be ranked by mass without redshift
information.  Mass properties will be explored in a future paper; here
we wish to describe the optical and X-ray counterparts and
spectroscopic followup of the cluster candidates.  With spectroscopic
redshifts in hand, a future paper will deal with lensing masses, X-ray
luminosities, and other redshift-dependent quantities.  However, to
set a rough mass scale to guide the reader's expectations, we note
that the mass of Candidate 8 has already been given by Wittman \etal\
(2003) as $8.6 \pm 2.3 \times 10^{14}\ (r/1\, {\rm Mpc})\ M_\odot$
within radius $r$, assuming a singular isothermal sphere profile.
This candidate does not show significantly more shear than the
lowest-ranked candidate, so it may be taken as an rough guide to the
mass threshold at its redshift, $z=0.68$.  This threshold changes with
redshift in a way that requires knowing the source redshift
distribution in detail, but as a rough guide, it is expected to fall
by a factor of nearly two by redshift 0.35, and then rise again toward
lower redshift.  Therefore one would not expect to find a cluster in
this sample with mass much less than $\sim 4\times 10^{14}\ (r/1\,
{\rm Mpc})\ M_\odot$, although we caution that this statement is
model-dependent.  The same cluster provides a rough guide to the
significance threshold in the current sample: roughly 4$\sigma$ based
on the errors quoted in that paper.

Finally, we required a splitting criterion before making the final
ranking presented in Table~\ref{table-candidates}.  That is, at what
angular separation would a secondary peak be counted as a candidate in
its own right, rather than as part of a higher-ranked candidate?
Because redshift information was not available at the time, angular
separation was the only criterion available.  We based the decision on
the practicality of X-ray followup.  If several clumps could fit
comfortably within the available field (formally, 8\arcm\ separation
or less), they were considered a single candidate; otherwise, they
were split.  This definition of a cluster is rather frugal; the
angular resolution of the convergence maps is $\sim$2\arcm, and a
definition based on that angular scale would have yielded more
candidates above a given shear threshold.  Still, 8\arcm\ corresponds
to 3.2 Mpc (comoving) at a typical lens redshift of 0.4, so there is
some justification for considering such multiple clumps to be
physically associated if they are at the same redshift.  When
shear-selected samples are compared with n-body simulations to
constrain cosmological parameters, the exact splitting criteria will
matter less than the fact that the same criteria can be applied to
observations and simulations without bias.

Many of the steps in this procedure have adjustable parameters, and
they may be far from optimized.  We would like to explore different
algorithms for making convergence maps, or fitting the shear field
directly for cluster models.  In additon, photometric redshift
information has become available since the initial selection, allowing
for the possibility of a tomographic filter for the cluster selection.
Future papers will explore these issues, as well as rigorously define
a complete sample.  The focus of this paper is the sample as defined
here, which was frozen rather early in the life of the DLS to
accommodate the logistics of the followup.

\subsection{Followup Program}

We pursued a multiwavelength followup program with these components:

\begin{itemize}
\item Literature search for known clusters.  We searched the NASA/IPAC
Extragalactic Database (NED) for known clusters within 5\arcm.  This
generally provided little information, as most of the sky has not been
searched deeply for clusters.  Any overlap with a deep optical or
X-ray survey would be fortuitous, so only bright, low-redshift
clusters from all-sky surveys would be expected.  There was only one
unambiguous identification from NED, of the top-ranked candidate which
happens to be Abell 781.

\item X-ray spectro-imaging with the \CXO\ X-ray Observatory, detailed
  in the next subsection.

\item Inspection of multicolor images, to find clustered galaxies
  (clustered both spatially and in color) and/or lensed arcs.  One
  candidate had an obvious arc and has already been published (Wittman
  \etal\ 2003).  We did not attempt a quantitative definition of
  galaxy overdensity, partly because the color information became
  available only gradually.  Now that color information is available
  for all clusters, an objective optical search is underway.  Based on
  our subjective judgement, all candidates were found to be associated
  closely enough with clustered galaxies to justify spectroscopic
  followup.

\item Literature (NED) search for any galaxies within 5\arcm\ with
spectrocopic redshifts, to make the spectroscopic followup more
efficient.  Specifically, field F4 overlaps the 2dF Galaxy Redshift
Survey (2dFGRS catalog; Colless \etal\ 2001), and at several points in
this paper we will refer to that survey.  However, the 2dFGRS
generally doesn't go deep enough to identify redshifts of DLS
clusters.  With one exception, its role was limited to identifying
foreground groups or providing 1---2 extra member redshifts.

\item Spectroscopic redshifts.  These are necessary because
  photometric redshifts are not precise enough to conclusively rule
  out line-of-sight projections.  At the same time, lack of a tight
  cluster red sequence is not strong evidence for a projection,
  because many real clusters have weak red sequences.  We observed
  with the Low-Resolution Imaging Spectrograph (LRIS, Oke \etal\ 1995)
  on the Keck I telescope on several runs: November 2000 (DLSCL
  J1055-0503), December 2003 (DLSCL J0916.0+2931, DLSCL J0916.0+3025)
  and April 2005 (DLSCL J1402.2-1028, DLSCL J1402.0-1019, DLSCL
  J1048.4-0411).  We also used the Hydra spectrograph on the CTIO 4-m
  Blanco telescope in March 2004, obtaining redshifts for DLSCL
  J1049.6-0417 and DLSCL J0522.2-4820.

\end{itemize}

\subsection{X-ray Data}
\label{subsec-xray}

X-ray imaging surveys, first with the {\it Einstein Observatory} and
subsequently with {\it ROSAT}, have yielded large samples of uniformly
selected galaxy clusters out to cosmologically interesting redshifts.
Such surveys have provided a wealth of information on the properties
of clusters and how those properties evolve with cosmic time (see
e.g., Rosati, Borgani \& Norman~2002).  Because the X-ray emission of
clusters depends on the square of the gas density, it is the
observable least susceptible to line-of-sight projections, which
provides an important motivation for our X-ray follow-up activities.
However, another advantage of X-ray follow-up is the potential it
offers us to relate the X-ray properties of our shear-selected
clusters to the existing large body of knowledge on X-ray--selected
clusters.  A first step in this direction, presenting the X-ray
luminosity--temperature relation of shear-selected clusters, is given
in Hughes et al.~(2005, in preparation).  In the following we describe
how we acquired and analyzed our X-ray followup data.

We chose pointed followup of our candidates rather than an X-ray
survey of the DLS area based on practical considerations.  A survey
would require a great deal of telescope time and would generally be
shallower than pointed observations. Pointed observations do not
reveal how many X-ray--selected clusters might have been missed by
shear selection, but that is not our primary interest here and indeed
the literature already contains a great deal of lensing followup of
X-ray--selected clusters to address that question.  We concluded that
the most efficient use of expensive satellite time was a pointed
followup program.

Note that high angular resolution is necessary for the zero
false-positive rate assumed for X-ray selection.  Older X-ray
facilities such as the Position Sensitive Proportional Counter
onboard {\it ROSAT} with its on-axis angular resolution of 25\arcs\
resulted in a 10\% false positive rate, primarily from blends of point
sources (Vikhlinin \etal\ 1998).  Archival X-ray observations, even if
deep enough, would not be suitable.

As part of this project, candidates 2--8 were observed with the {\it
Advanced CCD Imaging Spectrometer} imaging array (ACIS-I) aboard the
{\it Chandra X-ray Observatory}.  The four ACIS-I front-side
illuminated chips, as well as chip S2, were active, although for
cluster detection only the four ACIS-I chips were used.  This
corresponds to a 16\arcm\ (square) field of view.  As described above,
this determined how we split candidates with multiple clumps.

The \CXO\ data were taken in timed exposure mode and events were
telemetered in VFAINT format.  Our first ranked candidate had been
observed previously for 10 ks and its data were extracted from the
archive. The other pointings were all nominally 20 ks long: after all
data processing and filtering steps the individual exposure times
varied from 18522 s to 20309 s.  Identical reduction procedures were
applied to all data sets.  Background was reduced using VFAINT mode
information and light curves were inspected to reject data during
times of high rates.  The gain map was updated, corrections for charge
transfer inefficiency were applied, and events were filtered for
status, grade (retaining only grades 02346), and bad pixels.

In at least half of the cases, an X-ray cluster was evident even in
the raw data as an extended X-ray source near the position of the mass
cluster. For consistency and in order to optimize our search for low
level diffuse emission, we applied the following procedure to all
observations. The photon energy range was restricted to the 0.5--2 keV
band. X-ray point sources were identified and replaced with Possion
noise at the level given by the average number of counts per pixel
elsewhere in the image.  Because of our primary interest in extended
X-ray sources, we employed a very loose definition of point source and
replaced any source that had even as few as 2 counts per pixel in the
unblocked data.  The count image was convolved with a gaussian
smoothing kernel (with $\sigma = 10^{\prime\prime}$), and divided by
the exposure map (calculated for a photon energy of 1 keV).  In all
but one case, one or more extended X-ray sources appeared in the final
processed image.  The full extent of each source was estimated from
the smoothed image and flux determination was done using this size.
We present the complete list of extended X-ray sources following the
description of each candidate.

\section{Cluster Descriptions}
\label{sec-cand}

In this section we present details on each cluster candidate in the
\CXO\ sample, including redshift and optical and X-ray
counterparts. In ranked order:

\subsection{DLSCL J0920.1+3029 (Abell 781)}

\begin{figure}
\centerline{\resizebox{4.5in}{!}{\includegraphics{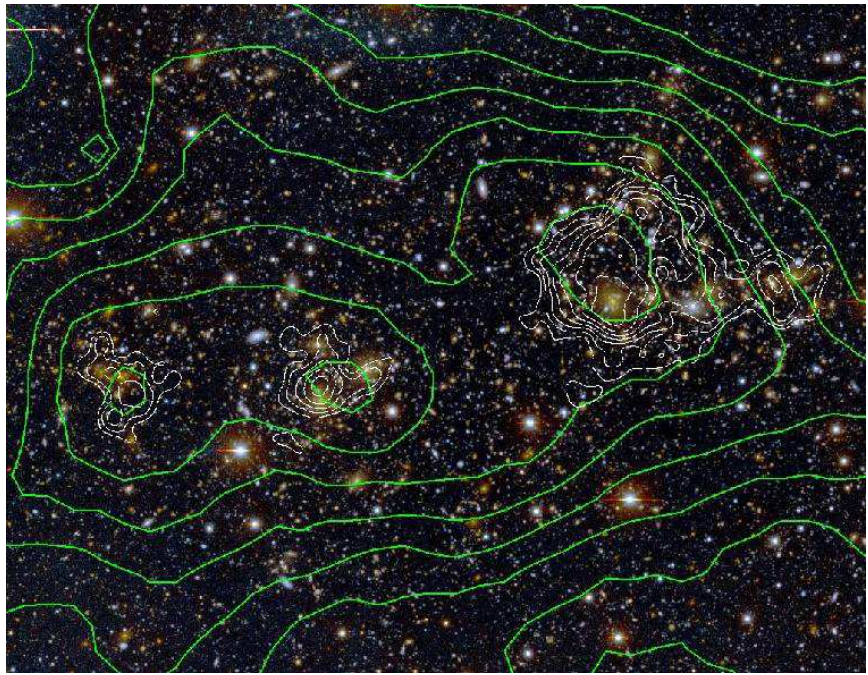}}}
\caption{DLSCL J0920.1+3029 (Abell 781): convergence map (green) and
  X-ray (white) contours overlaid on the multiband optical
  imaging. North is up, east left, and the field size shown is
  17\arcm\ diameter (4.5 Mpc at $z=0.298$). Clumps are referred to in the text
  as A, B, C, from west to east. For all figures, point sources have
  been removed from the X-ray data.\label{fig-c1}}
\end{figure}

The most prominent shear-selected cluster in the sample is an Abell
cluster, Abell 781, listed at $z=0.298$ by Struble \& Rood (1999) and
at $z=0.295$ by Bohringer \etal\ (2000).  This is a fairly rich
cluster (Abell richness class 2).  Three separate clumps are resolved
in the convergence map, with a maximum separation of 10\arcm.  Because
these all fit within the ACIS-I field of view, they were counted as a
single candidate.  Figure~\ref{fig-c1} shows the convergence contours
(green) overlaid on a BVR color composite image.  We propose to name
these clumps A, B, and C, from west to east (also in decreasing order
of galaxy richness and X-ray flux).  The position for Abell 781 given
by Abell, Corwin \& Olowin (1989; hereafter ACO) coincides most
closely to clump A, but differs by $\sim$3\arcm\ in declination (the
positional error quoted by ACO is 2.5\arcm).  Clumps A and B are
identified separately in the Northern Sky Optical Cluster Survey (Gal
\etal\ 2003) .

\clearpage\clearpage

Clump A was detected in the X-ray band in the ROSAT All-Sky Survey
catalog (Voges \etal\ 1999).  \CXO\ observed this cluster on 3 October
2000 (Obsid \#534) and in Figure~\ref{fig-c1} we plot the contours of
X-ray emission from these data in white.  (In all cluster figures,
unless otherwise noted, X-ray contours are shown after
removing point sources and smoothing).  All three lensing peaks appear
as extended X-ray sources, with very good positional matches to the
shear peaks.  In addition, with the good angular resolution of \CXO,
clump A can be split into a main clump (to which we continue to refer
as A) and a new clump designated D, centered 2\arcm\ to the west of
clump A's center.

The multiband optical imaging shows a cluster of galaxies at each
clump, with very good positional and morphological agreement among
galaxies, X-ray, and lensing (although in optical and lensing, the
split between clumps A and D is not resolved).  In addition, the order
of galaxy richness closely follow the order of descending X-ray flux
from A to C.  However, it is clear from the colors of the galaxies
that clump C is at a redshift $\sim$0.1 higher than clumps A and B.
We are pursuing confirmatory spectroscopy of this clump.

In summary, Abell 781 proper consists of two main clumps separated by
6.5\arcm, or 1.75 Mpc transverse at $z=0.298$. One of these clumps is
resolved into two subclumps in X-ray only.
Another cluster at $z\sim 0.4$ appears 4\arcm\ east of the eastern
clump of A781 proper.  Each of these three clumps is detected in
lensing, X-ray, and galaxies.

A detailed paper comparing the lensing and X-ray morphologies of all
clumps is in preparation.  We defer the question of whether this
structure should count as one, two, or three shear-selected clusters
to Section~\ref{sec-summary}.

\subsection{DLSCL J0522.2-4820}

\begin{figure}
\centerline{\resizebox{4.5in}{!}{\includegraphics{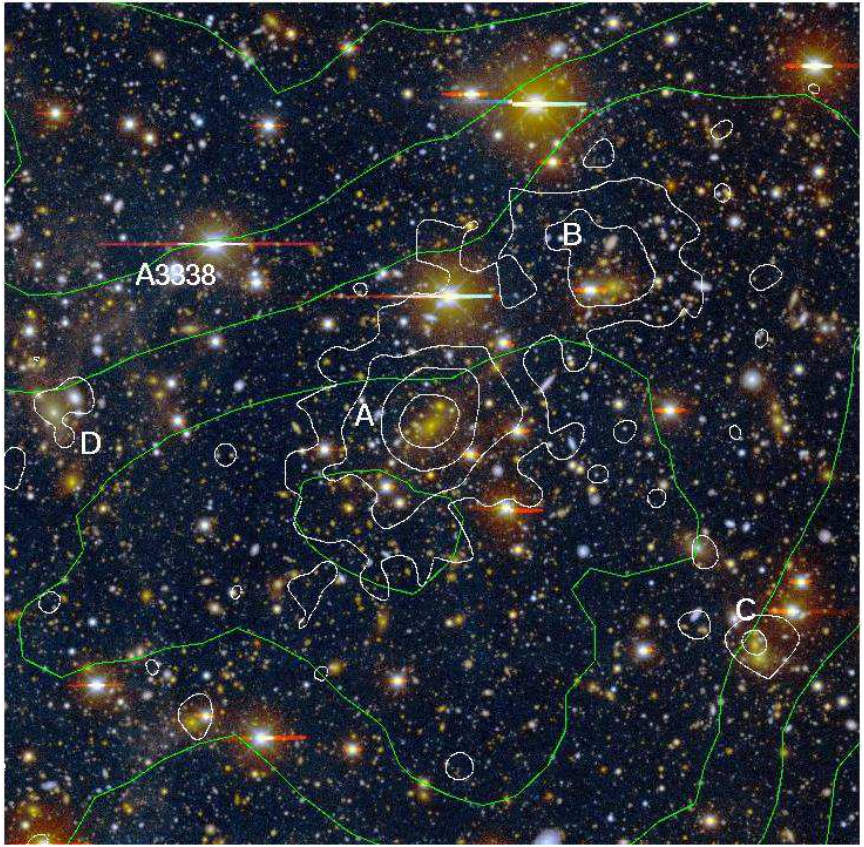}}}
\caption{DLSCL J0522.2-4820: convergence map (green) and X-ray (white)
  contours overlaid on the multiband optical imaging. North is up,
  east left, and the field size shown is 12\arcm\ diameter (3.2 Mpc
  transverse at $z=0.296$). Abell 3338 is more likely associated with
  the lower-redshift group D than with DLSCL
  J0522.2-4820 (labeled A).\label{fig-c2}}
\end{figure}

The second-ranked candidate is 5.2\arcm\ from an already-known
cluster, Abell 3338, but, based on the evidence we develop below, is
probably not the same structure identified by ACO.  NED does not
contain any other clusters or galaxies with known redshift in the
area.  Figure~\ref{fig-c2} shows the convergence contours (green)
overlaid on the optical imaging.  Also marked is the nominal position
of Abell 3338.

\clearpage\clearpage

We obtained X-ray spectro-imaging of this cluster with \CXO\ on 27
June 2003 (Obsid \#4208).  The X-ray contours are shown in white in
Figure~\ref{fig-c2}.  A luminous extended X-ray source appears at
05:22:15.6 -48:18:17, $\sim$2\arcm north of the convergence map peak.
A future paper will examine the significance and implications of
offsets between positions of X-ray, convergence, and galaxy locations.
Here, we are concerned with them only insofar as making a secure
cross-identification.  As a guide to the uncertainty in the current
lensing positions, we note that the position of a different cluster,
DLSCL J1054.1-0549, changed by $\sim$1\arcm\ when additional $R$ data
became available and we restacked its subfield.  Therefore, a 2\arcm\
displacement between the X-ray and lensing positions of this cluster
is not large enough to cast significant doubt on the
cross-identification.

A second, fainter, extended X-ray source is centered at 05:21:59.6
-48:16:06, 3\arcm\ northwest of the primary source.  There is no
convergence map peak at this position, but there is a definite
extension in this direction from the main cluster.  A third extended
X-ray source appears at 05:21:47.6 -48:21:24, or 5.5\arcm\ southwest
of the main source, and a fourth at 05:22:46.6 -48:18:04, or 5.2\arcm\
east of the main source.  There are no convergence map peaks at these
positions, nor extensions toward the positions.  We label the four
X-ray sources A, B, C, and D, in decreasing order of X-ray flux.

All four extended X-ray sources are coincident with clustered
galaxies, with the galaxy richness generally tracking the X-ray
brightness.  There is no measurable positional offset between the
X-ray positions and the galaxy clumps.

In March 2004 we obtained redshifts of 28 galaxies in the area around
the main cluster with the Hydra instrument on the CTIO 4-m Blanco
telescope.  
Sixteen members of clump A were identified with a mean redshift of
$0.296 \pm 0.001$.  Three members in clump D were identified at
$z=0.21$.  No spectroscopy is available for clumps B and C, but the
photometry suggests that they are at the same redshift as the main
cluster, and are clearly at higher redshift than clump D.

Clump D is closer than any of the aforementioned positions to the
nominal position of Abell 3338, with an offset of 2.5\arcm, equal in
size to the positional error quoted by ACO.  Clump D is also at lower
redshift and its galaxies have higher surface brightness than those of
the other clumps.  Therefore, if only one clump was detected by ACO,
it should have been clump D.  Furthermore, it seems unlikely that ACO
conflated multiple clumps, because their position is not near the mean
position of any set of clumps.  Thus we suggest that clump D, not the
main lensing/X-ray clump A, is Abell 3338.  ACO list the redshift as
0.045, but they note that 0.045 is inconsistent with the redshift
expected from the magnitudes of the cluster members.  Furthermore, the
redshift source is given only as a private communication, and there is
no indication of how many galaxies it is based on.  Therefore we
believe the true redshift of Abell 3338 (clump D) is 0.21, not 0.045.

In summary, the lensing peak is coincident with a bright, extended
X-ray source and a cluster of galaxies at $z=0.296$.  Two secondary
X-ray sources and galaxy clumps are possibly at the same redshift and
simply subclumps of the main cluster.  If so, they are not massive
enough to cause a lensing signal, except perhaps for an extension in
the direction of clump B.  In addition, a cluster at lower redshift
($z=0.21$) and 5.2\arcm\ to the east of the main cluster, appears to
be Abell 3338, which is detected in X-ray and galaxies but not on the
current convergence map.  Note that Abell 3338 is listed as a richness
class 0 cluster.  According to Briel \& Henry (1993), richness class 0
Abell clusters are on the average a factor of two (four) less X-ray
luminous than richness class 1 (2) Abell clusters, but with
considerable scatter.

\subsection{DLSCL J1049.6-0417}

\begin{figure}
\centerline{\resizebox{4.5in}{!}{\includegraphics{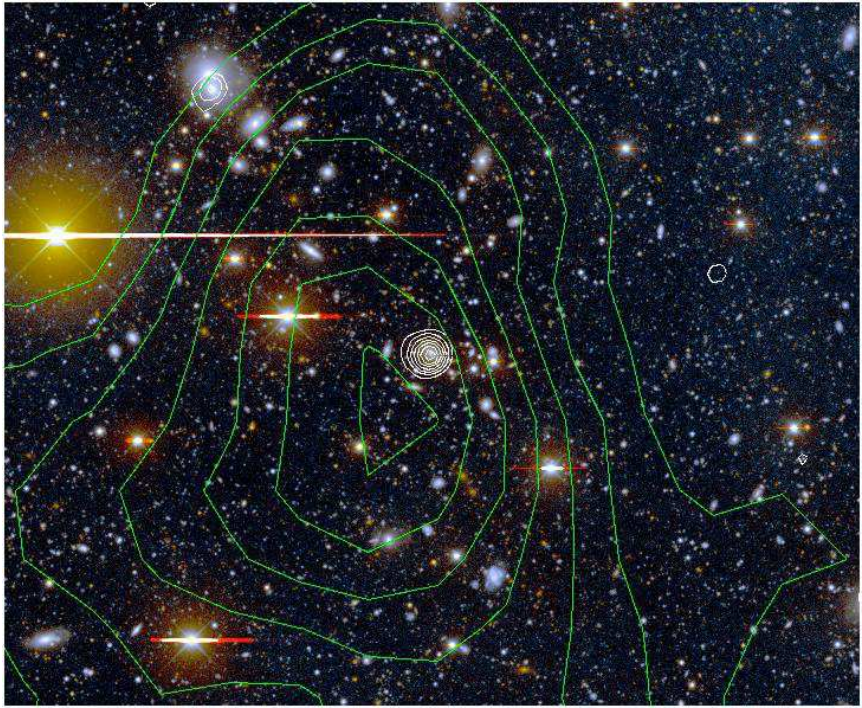}}}
\caption{DLSCL J1049.6-0417: convergence map contours (green) and
  X-ray contours (white) overlaid on the multiband optical
  imaging. North is up, east left, and the field size shown is
  10\arcm\ diameter (2.5 Mpc transverse at $z=0.267$).\label{fig-c3}}
\end{figure}

The third-ranked candidate has not been previously identified.
Figure~\ref{fig-c3} shows the convergence contours (green) overlaid on
the optical imaging.

\clearpage\clearpage

The X-ray contours from \CXO\ spectro-imaging obtained on 2 March 2003
(Obsid \#4210) are shown in white in Figure~\ref{fig-c3}.  A slightly
extended X-ray source is centered at 10:49:37.9 -04:17:28, offset
41\arcs\ from the convergence peak.  The X-ray position is coincident
with a cluster of galaxies, with no measurable positional offset
between the X-ray centroid and the brightest cluster galaxy (BCG).
A second, fainter extended X-ray source is visible 5\arcm\ to the
northeast of the main cluster (10:49:50.7 -04:13:38), corresponding to
a group of very large, bright galaxies.

We obtained spectroscopy of this cluster in the March 2004 Hydra run,
obtaining redshifts of 19 galaxies.  The mean redshift of the cluster
is 0.267 $\pm 0.002$ based on seven members.  The redshift of the
second group and X-ray source to the northeast is $z=0.068$ based on
eight members listed in the 2dFGRS.  This is too low to significantly
affect the convergence map, and indeed there is no deviation of the
convergence contours toward this group.

\subsection{DLSCL J1054.1-0549}

\begin{figure}
\centerline{\resizebox{4.5in}{!}{\includegraphics{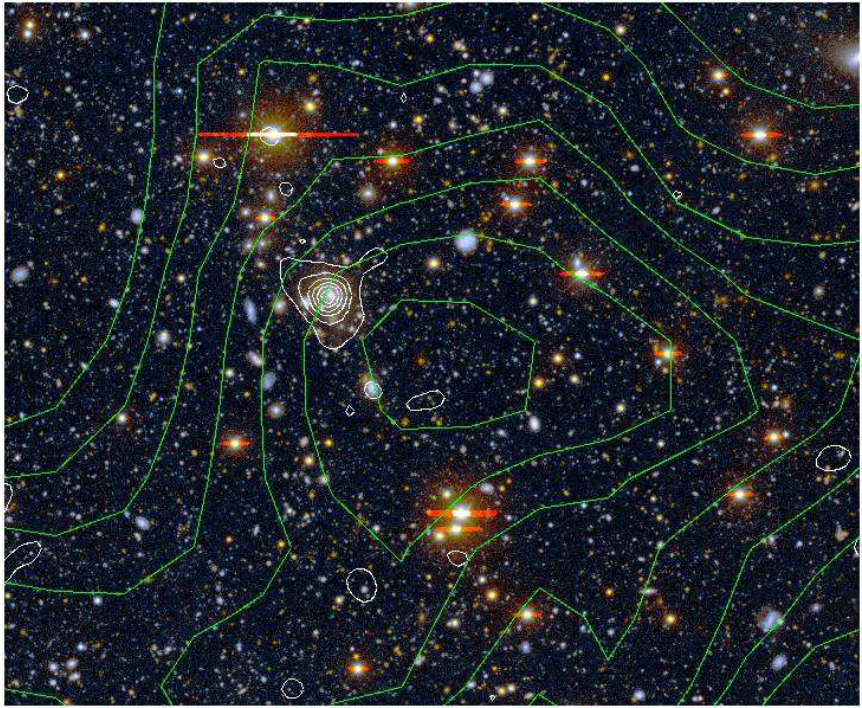}}}
\caption{DLSCL J1054.1-0549: convergence map (green) and
  X-ray contours (white) overlaid on the multiband optical
  imaging. North is up, east left, and the field size shown is
  10\arcm\ diameter (1.9 Mpc transverse at $z=0.190$).\label{fig-c4}}
\end{figure}

The fourth-ranked candidate, shown in Figure~\ref{fig-c4}, has not
been previously identified in the literature.  \CXO\ imaging from 3
March 2003 (Obsid \#4211) reveals a single luminous extended source
centered at 10:54:14.8 -05:48:50, offset $\sim$2\arcm\ from the
convergence peak.  The X-ray position is coincident with what appears
to be the BCG of a cluster of galaxies.  Several member spectroscopic
redshifts are available from the 2dFGRS.  That database lists five
probable members with a mean redshift of $0.190 \pm 0.001$, which we
adopt for this paper.

\clearpage\clearpage

\subsection{DLSCL J1402.2-1028}

\begin{figure}
\centerline{\resizebox{4.5in}{!}{\includegraphics{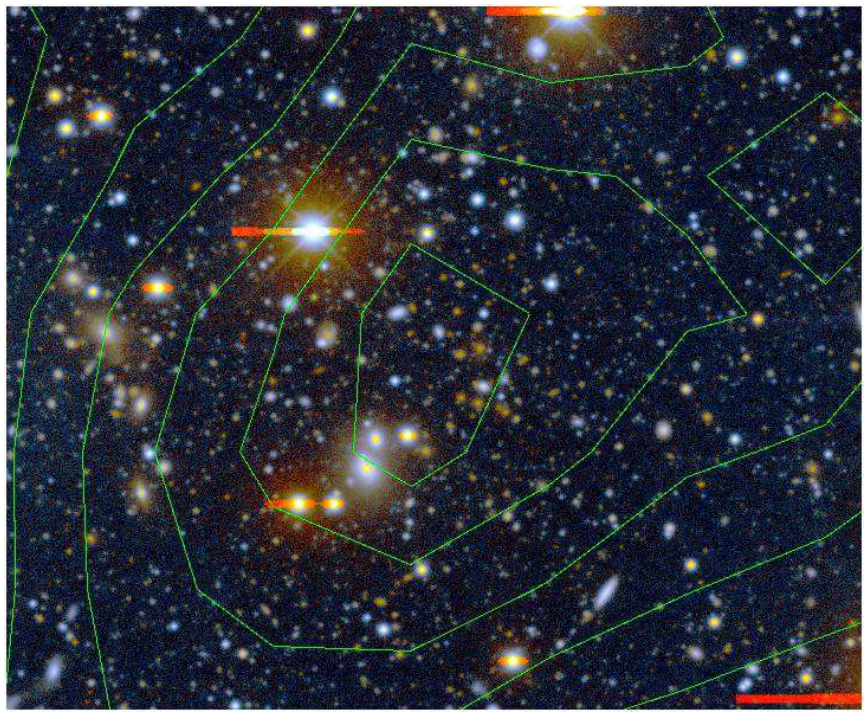}}}
\caption{DLSCL J1402.2-1028: convergence map contours (green) overlaid
  on the multiband optical imaging. X-ray contours are omitted because
  there is no significant X-ray source. North is up, east left, and
  the field size shown is now 5\arcm\ diameter to better show the
  details described in the text.\label{fig-c5}}
\end{figure}

The fifth-ranked candidate (Figure~\ref{fig-c5}) is perhaps the most
interesting.  There are no previously-known clusters, or even galaxies
with known redshifts, in the literature within 5\arcm\ of this
location.  The \CXO\ data from 19 March 2003 (ObsId \#4213) reveal no
significant X-ray source, with a 90\% confidence upper limit of
8$\times$10$^{-15}$ erg cm$^{-2}$ s$^{-1}$ in the 0.5--2 keV band.
However, the lensing position coincides with a modest group of red
galaxies.  Lacking a dominant BCG, the group position is not
measurable to great accuracy, but its centroid is no more than
30\arcs\ from the lensing position.

\clearpage\clearpage

There is also a group of lower-redshift (based on angular size,
magnitude and color) galaxies 2.4\arcm\ to the east of the convergence
map peak.  At that position, the convergence map value has fallen to
less than half its peak value.  In contrast, the offsets of the
previous clusters, although approaching 2\arcm, did not involve a
large drop in the convergence map value at their position.  Based on
that, we found the most likely cross-identification of the lensing peak
to be with the higher-redshift group.

Still, with no detected X-ray emission, a line-of-sight projection is
naturally suspected.  Photometric redshifts indicated $z\sim0.3$ for
the low-redshift group, and $z\sim0.5$ for the high-redshift group, so
we designed one LRIS slitmask for low redshifts and another for higher
redshifts.  The low-redshift slitmask received less exposure time, and
was shifted slightly in position, but there was a great deal of
overlap in area.  This allowed us to observe some galaxies through the
``wrong'' slitmask if the constraints on slit position in the more
appropriate slitmask did not work out.  There were 16 slits in the
low-redshift slitmask and 19 in the high-redshift one.

\begin{figure}
\centerline{\resizebox{3in}{!}{\includegraphics{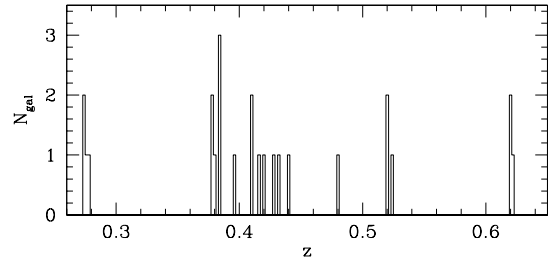}}}
\caption{DLSCL J1402.2-1028: redshift distribution.  The lack of a peak,
  in combination with lack of X-ray emission, points to this candidate
  being a projection.\label{fig-c5zdist}}
\end{figure}

We observed on 11 April 2005, and obtained 29 secure redshifts.  The
redshift distribution, shown in Figure~\ref{fig-c5zdist}, shows no
peak.  The precision of the redshifts in this dataset, based on the
error in the mean of typically $\sim$6 lines, is typically
$\sim$0.0005, and the bins plotted are 0.002 wide.  Therefore a
cluster should occupy $\sim$3 contiguous bins.  Instead, pairs and
triplets of galaxies are seen at many different redshifts.  If this is
a projection, it is of numerous nearly unrelated galaxies rather than
of two readily identifiable groups.

\clearpage\clearpage

The current redshift data do not conclusively prove that there is no
cluster.  It is possible that there is a cluster at $z=0.66$ (or
higher) and the targeting simply did not go deep enough to obtain many
members.  This seems somewhat unlikely, given that we obtained
redshifts of four of the eight red galaxies in the central 50\arcs\
bright enough for spectroscopy, and no two lie at the same
redshift. Only one of these is at $z=0.66$.  A second possibility is
that the lower-redshift group 2.4\arcm\ to the east is responsible for
the lensing peak despite its offset.  This group has a
spectroscopic redshift of 0.28 and does not appear as a large peak in
Figure~\ref{fig-c5zdist} simply because it was not considered an
important target.  However, on balance the evidence for either of
these scenarios is weak.

\subsection{DLSCL J1402.0-1019}

\begin{figure}
\centerline{\resizebox{4.5in}{!}{\includegraphics{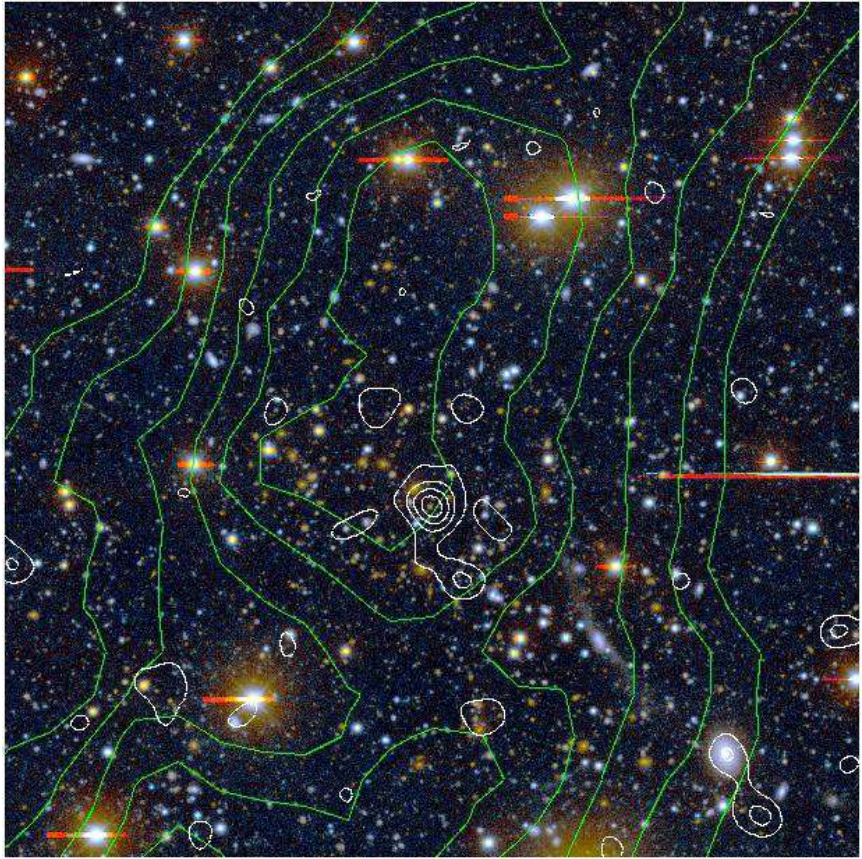}}}
\caption{DLSCL J1402.0-1019: convergence map contours (green) and
  X-ray contours (white) overlaid on the multiband optical
  imaging. North is up, east left, and the field size shown is
  10\arcm\ diameter (3.4 Mpc transverse at $z=0.427$).\label{fig-c6}}
\end{figure}

This sixth-ranked candidate (Figure~\ref{fig-c6}) consists of a
prominent ridge in the convergence map with two small summits of
nearly equal height on top of the ridge, one at 14:02:01 -10:19:35 and
the other at 14:02:02 -10:22:44.  Because these peaks are separated by
much less than one ACIS-I field size, they were considered one
candidate rather than two.  At a lower level the ridge in the
convergence map extends further south, to DLSCL J1402.2-1028.
However, there is a much more clear separation between DLSCL
J1402.2-1028 and DLSCL J1402.0-1019 than between the two peaks
comprising DLSCL J1402.0-1019.  There are no previously-known
clusters, or even galaxies with known redshifts, in the literature
near this location.

\clearpage\clearpage

The \CXO\ X-ray data taken 3 Sept 2003 (ObsId \#4214) reveal an
extended source centered at 14:01:59.7 -10:23:01.5, or 39\arcs\ from
the lensing position of the southern peak.  There is no detected X-ray
flux at the position of the northern peak.
 
The multiband optical imaging shows a modest cluster of galaxies at
the southern position.  This cluster lacks a dominant BCG and is
fairly amorphous.  Of the two brightest galaxies, the one at
14:02:00.54 -10:22:49.20 is somewhat more centrally located, so we
take that as equivalent to the BCG position. This is 22\arcs\ from the
lensing position and 18\arcs\ from the X-ray position.

We obtained Keck/LRIS redshifts of 18 galaxies in the region on 11
April 2005.  We identified ten members with a mean redshift of 0.4269
$\pm$ 0.0005.

\subsection{DLSCL J0916.0+2931}

\begin{figure}
\centerline{\resizebox{4.5in}{!}{\includegraphics{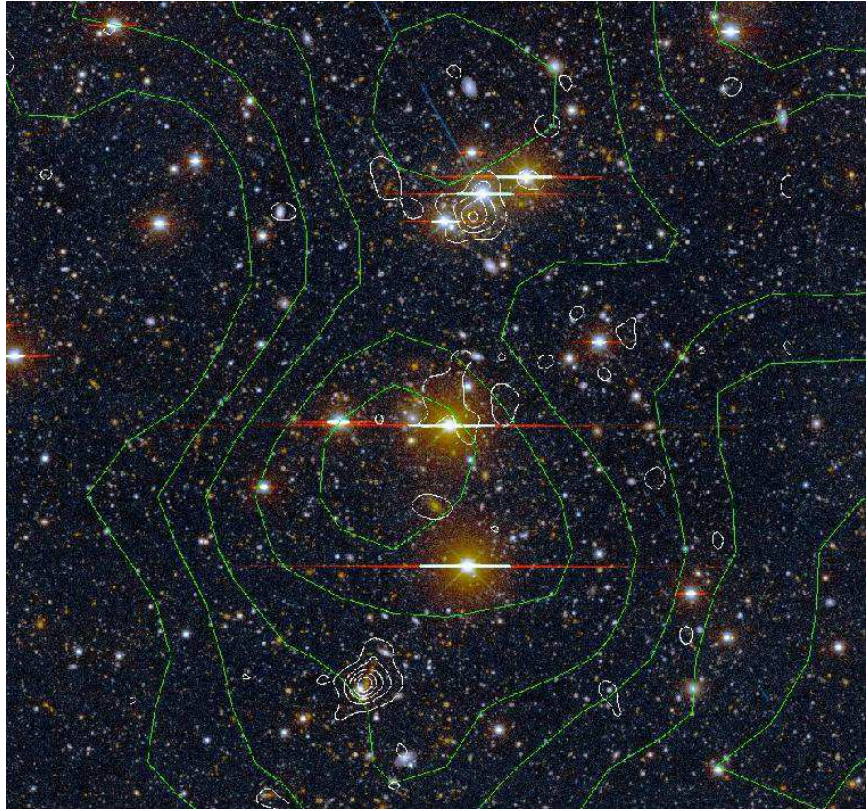}}}
\caption{DLSCL J0916.0+2931: convergence map (green) and X-ray (white)
  contours overlaid on the multiband optical imaging. North is up,
  east left, and the field size shown is now 15\arcm\ diameter (5.7
  Mpc transverse at $z=0.531$).
\label{fig-c7}}
\end{figure}

This candidate, on the convergence map, consists of a peak on top of a
north-south ridge (Figure~\ref{fig-c7}).  The ridge extends
$\sim$5\arcm\ to the south, and much further north.  There is a small
local maximum on the ridge $\sim$7\arcm\ to the north, but this bump
would not have qualified as a separate candidate even had it been
outside the ACIS-I field of view centered on the main peak.  The
literature contains no cluster or galaxy of known redshift in this
region.
\clearpage

\clearpage\begin{figure}
\centerline{\resizebox{2.5in}{!}{\includegraphics{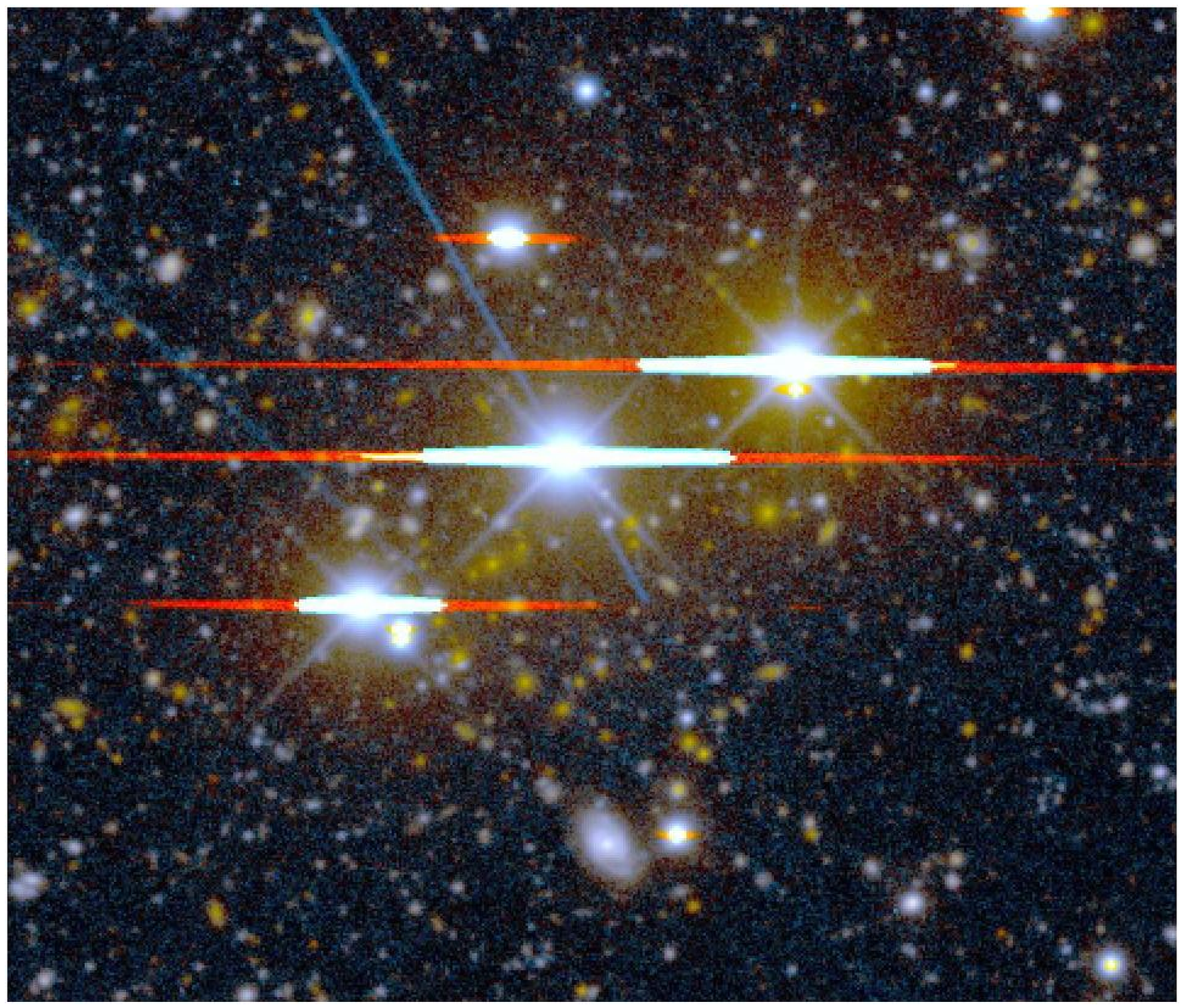}}}
\centerline{\resizebox{2.5in}{!}{\includegraphics{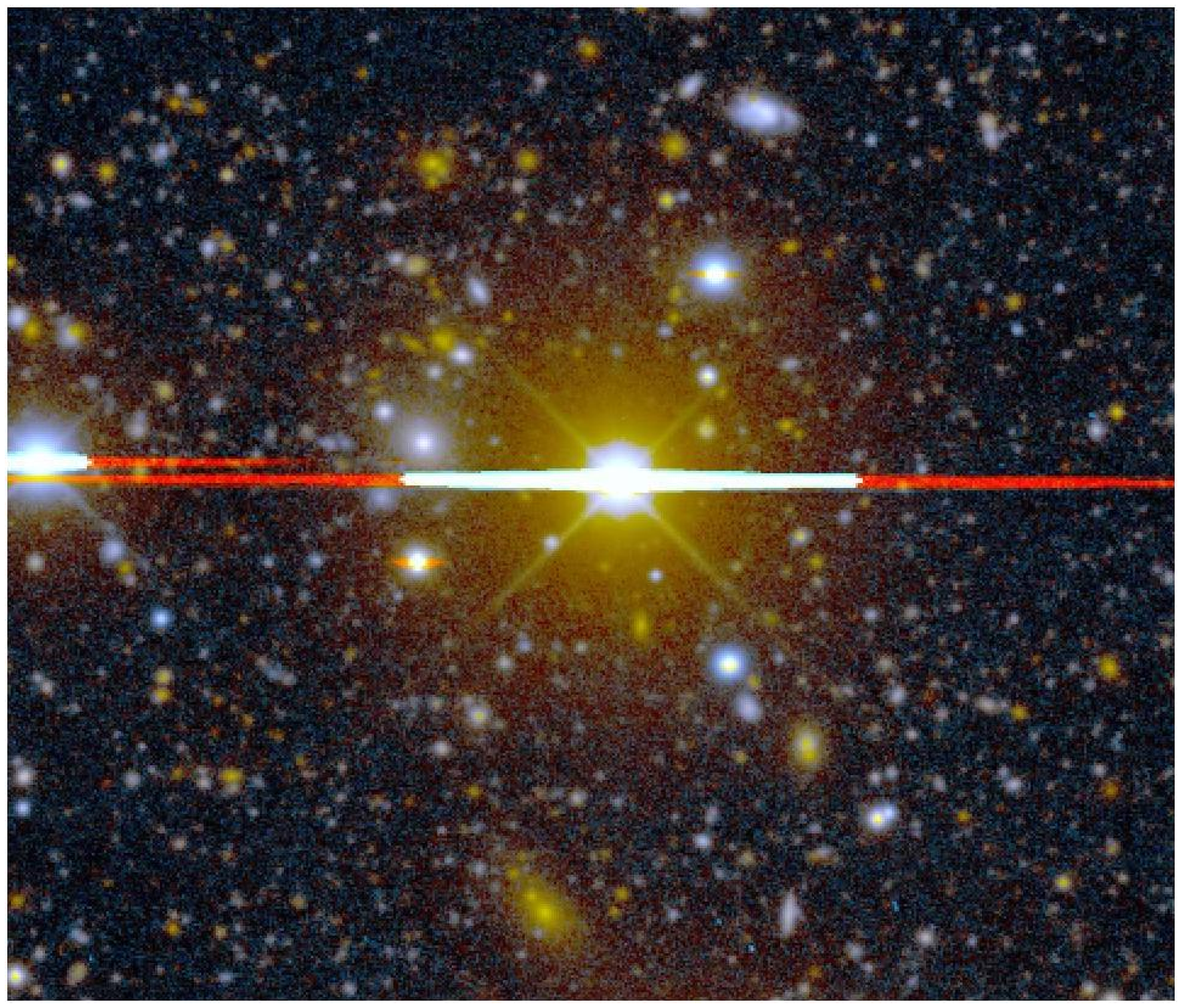}}}
\centerline{\resizebox{2.5in}{!}{\includegraphics{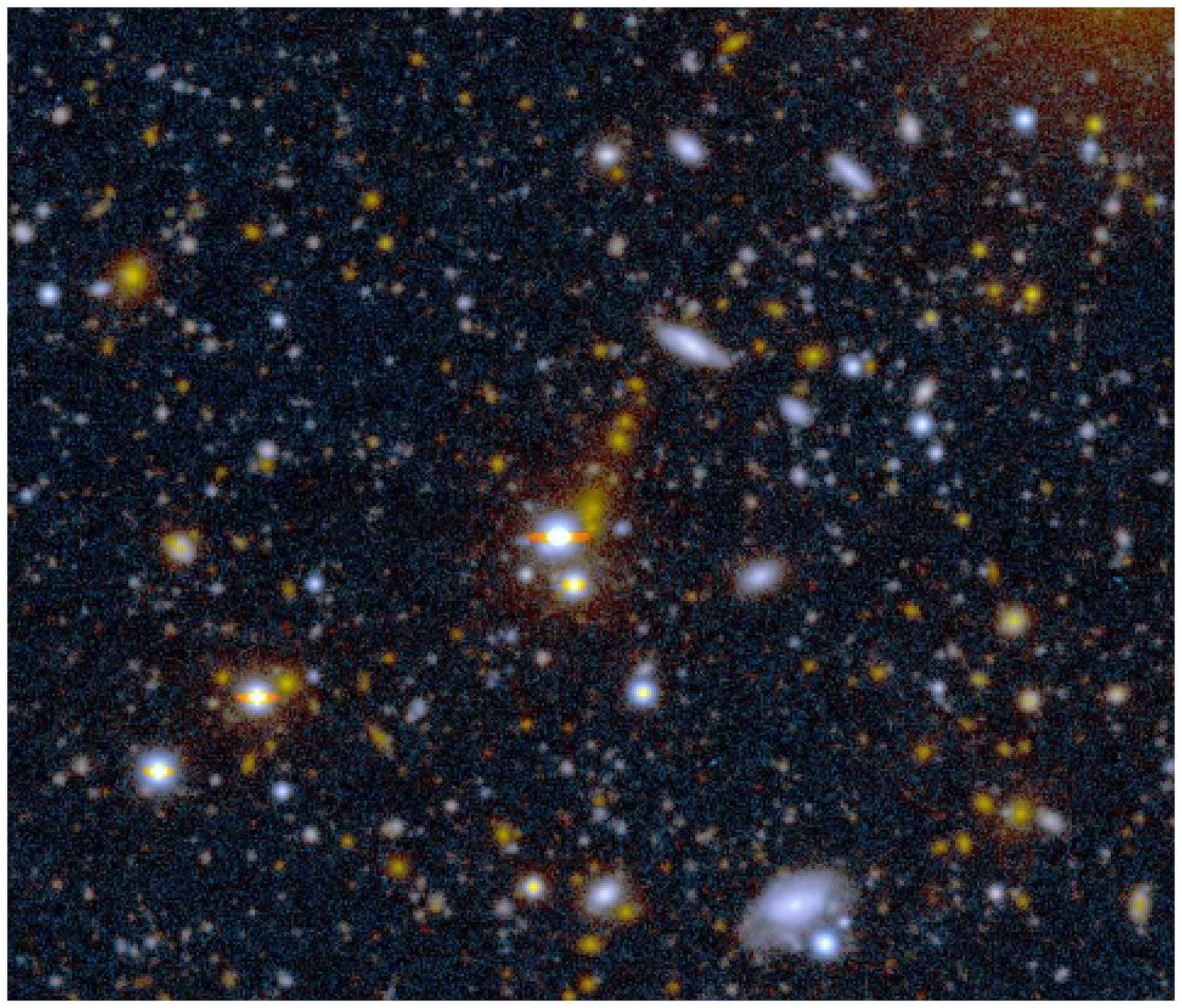}}}
\caption{DLSCL J0916.0+2931: details of galaxy clumps corresponding to
  the northern X-ray source (top), the central X-ray source and
  convergence peak (middle) and the southern X-ray source (bottom).
  The field size is 3.5\arcm\ (1.3 Mpc) diameter in each
  case. \label{fig-c7a}}
\end{figure}\clearpage

The \CXO\ data obtained on 16 December 2002 (ObsId \#4209) reveal two
extended X-ray sources of equal brightness.  The northern one, at
09:15:51.8 +29:36:37, is on the ridge in the convergence map between
the main and secondary convergence peaks, but much closer (2.4\arcm)
to the latter.  The southern X-ray source, at 09:16:01.1 +29:27:50,
lies on the southern extension in the convergence map.  Fainter, more
extended emission is also seen near the central convergence peak
(offset 2.1\arcm), with a centroid at 09:15:54.4 +29:33:16.

Close to this central cluster we detect a relatively bright X-ray
point source ($F_X \sim 4\times 10^{-13}$ erg cm$^{-2}$ s$^{-1}$ in
the 2--10 keV band), which is positionally coincident with the BL Lac
object FBQS J091552.3+293324 (also designated B2 0912+29).  We were
careful to take account of the full extent of the \CXO\
point-spread-function when removing the emission from this point
source.

Both brighter X-ray sources coincide with galaxy concentrations.  In
each case the galaxy clumps are amorphous and lacking a dominant
BCG (Figure~\ref{fig-c7a}, top and bottom), but for the northern clump
we must add the caveat that three $R\sim 12$ foreground stars severely
hamper our view.  In each case there is no measurable offset between
the X-ray position and the galaxy clump position.  There is a slightly
less convincing concentration of red galaxies at the position of the
fainter X-ray source, near the main lensing peak.  This area, too, is
partially hidden by another bright star (Figure~\ref{fig-c7a},
middle).

We obtained spectroscopy with Keck/LRIS on 20 December 2003, and in
the southern clump found 13 members with a mean redshift of $z=0.5306
\pm 0.0008$.  For the northern clump, we also have a redshift of 0.53
based on two galaxies in a longslit exposure.  There is no
spectroscopy of the central clump as yet, but the galaxy colors and
angular sizes are consistent with the same redshift.

In summary, there are two clusters of galaxies at $z=0.53$, separated
by 9\arcm, or 3.4 Mpc transverse in our adopted cosmology.  In the
center of the two is a third clump likely to be at the same redshift.
The north-south separation of the three clumps is reflected in the
ridge in the convergence map, which is also oriented north-south and
is at the same RA.  The convergence peak coincides with the central
clump, but this does not automatically imply that the central clump is
most massive.  It may be an effect of the smoothing in the convergence
maps.

\subsection{DLSCL J1055.2-0503}

\begin{figure}
\centerline{\resizebox{4.5in}{!}{\includegraphics{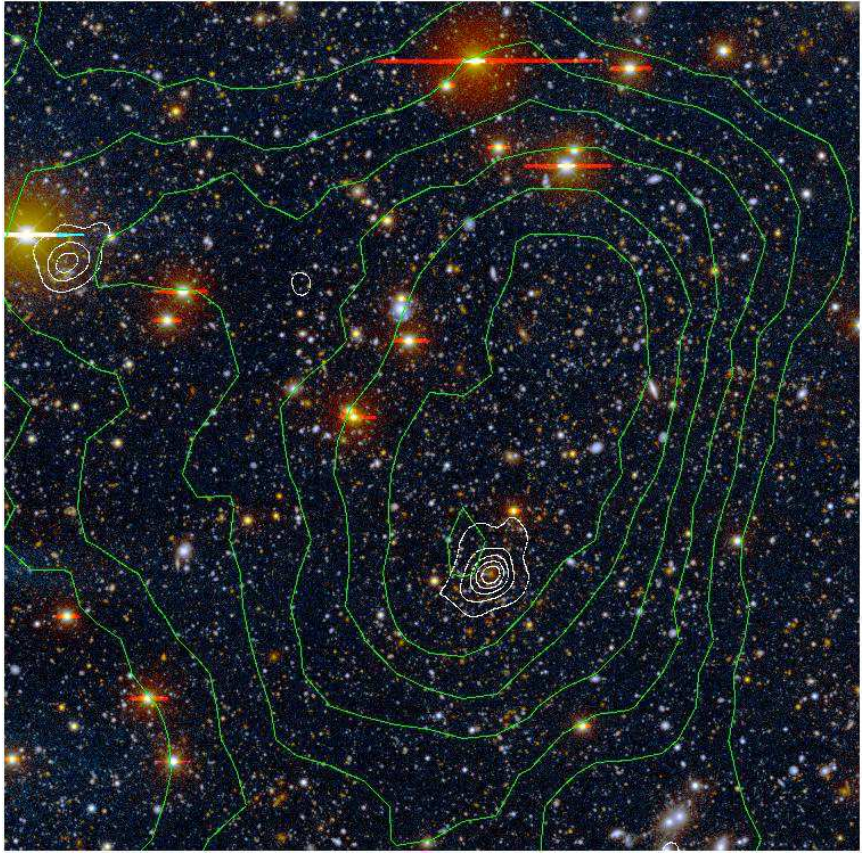}}}
\caption{DLSCL J1055.2-0503: convergence map contours overlaid on the multiband
  optical imaging. North is up, east left, and the field size shown is
  13\arcm\ diameter (5.6 Mpc transverse at $z=0.680$).
\label{fig-c8ovlay}}
\end{figure}

A description of this cluster (Figure~\ref{fig-c8ovlay}) has already
been published by the DLS (Wittman \etal\ 2003), although that paper
does not include the X-ray observations presented here.  It had not
been previously cataloged in the literature.

\clearpage\clearpage

The \CXO\ data from 8 March 2003 (ObsId \#4212) show an extended
source at 10:55:10.1 -05:04:14, or 43\arcs\ from the lensing position.
A secondary extended X-ray source appears at 10:55:35.6 -04:59:31, or
7.2\arcm\ northeast of the main cluster.  This does not correspond to
a lensing peak, although there appears to be some extension of the
lensing contours in that direction.

\begin{figure}
\centerline{\resizebox{2.5in}{!}{\includegraphics{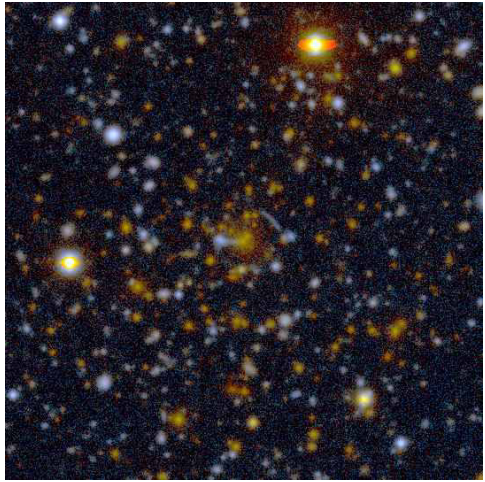}}}
\caption{DLSCL J1055.2-0503: detail of the central 2.3\arcm\ by
  1.6\arcm\ (1.0 by 0.7 Mpc).\label{fig-c8detail}}
\end{figure}

The multiband optical imaging shows a compact red cluster
with BCG position coincident with the main X-ray position.  There also
appear to be strong lensing features: A large blue tangential arc, and
an opposing blue radial arc (Figure~\ref{fig-c8detail}).  However, we
do not have redshifts of the arcs.  At the position of the second
X-ray source, the optical imaging shows a concentration of red
galaxies, but scattered light from a nearby bright star prevents
determination of a BCG or cluster morphology.

\clearpage\clearpage

We obtained Keck spectroscopy of the main cluster in November 2000,
and found a mean redshift of $z=0.680 \pm 0.001$ based on eleven
members.  We also obtained redshifts of the two galaxies closest to
the second extended X-ray source in April 2005, and found both to be at
$z=0.609$.  

\section{Summary and Discussion}
\label{sec-summary}

\subsection{Cross-identification with Previously Known Clusters}

Most of these clusters were not previously known.  The only
unambiguous case exception is the highest-shear cluster in the survey,
Abell 781.  Even in that case, one of the three shear peaks---the
higher-redshift one---did not correspond to a previously known
cluster.  This shear peak would have qualified as a candidate in its
own right had we used a smaller splitting radius.

A more ambiguous case is that of the second-ranked cluster, which is
near the listed position of Abell 3338.  The balance of evidence is
that the shear-selected cluster is not what ACO identifed as Abell
3338, but simply near it and at higher redshift.  The final case in
which some information existed in current databases is the
fourth-ranked cluster, DLSCL J1054.1-0549, which had enough 2dFGRS
redshifts to assign a cluster redshift, even though it hadn't been
identified as a group in any published work.

These facts are easily understood.  The redshift sensitivity of shear
selection is unlike that of any previous cluster selection technique.
It is insensitive at low redshift, whereas the vast majority of
clusters in current databases are precisely at low redshift because
all-sky surveys have been shallow.  Had the DLS overlapped with a deep
pencil-beam survey, no doubt most of these clusters would have been
identified by optical means.  For the same reason, a literature search
of known clusters inside the sample footprint, but missed by the shear
selection, is unlikely to yield interesting results.  Those clusters
are likely to have escaped shear selection simply by virture of being
too low-redshift.  Abell 3338 may be a counterexample, being
unselected at $z=0.21$, but there is some ambiguity regarding the
identification of Abell 3338, and in any case it may have escaped
shear selection because of low mass, as inferred from its low richness
(class 0).  For an adequate comparison of optical selection and shear
selection, the optical selection must be done on the DLS data itself,
which is the subject of a future paper (Margoniner \etal, in
preparation).

\subsection{Correspondence with Extended X-ray Sources}

Seven of the eight candidates exhibit detectable extended X-ray
emission.  The exception, DLSCL J1402.2-1028, is likely a
line-of-sight projection, judging by the spectroscopy.  In the other
seven cases, there are positional offsets of up to 2\arcm\ between the
X-ray sources and the shear-selected candidate positions, which may
not be significant in the current set of convergence maps.

In Table~\ref{table-xray} we present the candidate number, IAU
designation, position in epoch J2000, X-ray flux $F_X$, statistical
significance of detection (S/N), and redshifts of all the extended
X-ray sources in the eight X-ray data sets.  Fluxes are in the 0.5--2
keV band and have been corrected for Galactic absorption, which in all
cases was modest (in the range $N_{\rm H} \sim 2-4\times 10^{20}$
atoms cm$^{-2}$).  Where a candidate yielded multiple sources, the
sources are ordered by $F_X$.  

\begin{table}
\begin{center}
\begin{tabular}{|c|c|c|c|c|c|c|}
\hline
Candidate &IAU Designation& RA(J2000) & Dec(J2000) &
$F_X$\tablenotemark{a}&S/N & z (source)\tablenotemark{b}\\
\hline
1 & CXOU J092026+302938& 9:20:26.4 & 30:29:39 & $6.42\times 10^{-13}$ & 36.6 & 0.298 (1) \\
1 & CXOU J092053+302800& 9:20:53.0 & 30:28:00 & $1.16\times 10^{-13}$ & 13.3& 0.298 (1) \\
1 & CXOU J092110+302751& 9:21:10.3 & 30:27:52 & $9.47\times 10^{-14}$ & 12.2& $\sim$0.4 (2)\\ 
1 & CXOU J092011+302954& 9:20:11.1 & 30:29:55 & $4.98\times 10^{-14}$ & 9.1 & 0.298 (1) \\
\hline
2 & CXOU J052215-481816 & 05:22:15.6 & -48:18:17 & $2.6\times 10^{-13}$ & 30.2& 0.296 (3)\\
2 & CXOU J052159-481606& 05:21:59.6 & -48:16:06 & $5.9\times 10^{-14}$ & 10.4& $\sim$0.3 (2)\\
2 & CXOU J052147-482124& 05:21:47.6 & -48:21:24 & $8.3\times 10^{-15}$ & 4.9& $\sim$0.3 (2)\\ 
2 & CXOU J052246-481804 & 05:22:46.6 & -48:18:04  &  $4.7\times 10^{-15}$ & 3.3& 0.21 (5) \\ \hline
3 & CXOU J104937-041728& 10:49:37.9 & -04:17:29 & $1.07\times 10^{-14}$ & 5.3 & 0.267 (3) \\
3&CXOU J104950-041338& 10:49:50.7 & -04:13:38 & $5.6\times 10^{-15}$ & 3.7 & 0.068 (4)\\ \hline
4 &CXOU J105414-054849 & 10:54:14.8 & -05:48:50 & $1.32\times 10^{-14}$ & 6.7& 0.190 (4)\\ \hline
5\tablenotemark{c} & - & - & - & $<8\times 10^{-15}$ & - & - \\ \hline
6 &CXOU J140159-102301 & 14:01:59.7 & -10:23:02 & $6.6\times 10^{-15}$  & 4.1& 0.427 (5) \\ \hline
7 &CXOU J091551+293637 & 09:15:51.8 & 29:36:37 & $1.8\times 10^{-14}$ & 7.6& 0.53 (5) \\ 
7 &CXOU J091601+292750 & 09:16:01.1 & 29:27:50 & $1.8\times 10^{-14}$ & 7.4& 0.531 \\ 
7 &CXOU J091554+293316 & 09:15:54.4 & 29:33:16 & $7.09\times 10^{-15}$ & 4.6 & $\sim$0.5 (2) \\ \hline 
8 & CXOU J105510-050414& 10:55:10.1 & -05:04:14 & $2.21\times 10^{-14}$ & 8.7& 0.680 (6) \\ 
8 & CXOU J105535-045930& 10:55:35.6 & -04:59:31 & $1.58\times 10^{-14}$ & 7.5& 0.609 (5) \\ \hline
\end{tabular}
\end{center}
\caption{Extended X-ray sources in \CXO\ followup.\label{table-xray}}
\tablenotetext{a}{ erg cm$^{-2}$ s$^{-1}$ in the 0.5-2 keV band}
\tablenotetext{b}{Redshift sources: (1) Struble \& Rood (1999); (2)
  this paper, photometric; (3) this paper: CTIO 4-m/Hydra; 
(4) Colless  \etal\ (2001); (5) this paper: Keck/LRIS; (6) Wittman
  \etal\ (2003)}
\tablenotetext{c}{Upper limit at 90\% confidence.}
\end{table}

This is not to imply that each X-ray source listed directly
corresponds to the shear-selected candidate in its field.  The
relationships are not simple enough to encode in a table, so a quick
recap of each candidate with multiple X-ray sources is in order:

\begin{itemize}
\item[1.]{Each of the top three X-ray sources corresponds to its own shear
  peak.  The fourth source is a subclump of the top source, clearly
  split in the X-ray but not in convergence or galaxy distribution.}
\item[2.]{The top source corresponds to the shear-selected candidate,
  and the next two sources are likely to be subclumps of the same
  cluster, not resolved or detected separately in the current
  convergence maps.  The fourth source (Abell 3338) is unrelated.}
\item[3.]{The top source corresponds to the shear-selected candidate, and
  the second source is an unrelated low-redshift cluster with no
  visible effect on the convergence map.}
\item [7.]{The faintest source listed corresponds best to the position
of the convergence peak.  However, the morphology of the convergence
map indicates that the three subclumps were not well resolved and that
the peak position may simply be the center of the smoothed
distribution.}
\item [8.]{The top source corresponds to the shear-selected candidate.
The second source lies at a different redshift, but corresponds to an
extension in the convergence map contours.}
\end{itemize}

Extended X-ray sources with no apparent effect on their convergence
maps nevertheless correspond to galaxy overdensities in each
case. Excepting CXOU J104950-041338, all appear to be in the redshift
range appropriate for shear selection.  It is likely that they are
simply not massive enough to make it into the current sample.  This
will be verified, assuming some mass-$L_X$ relation, after analyzing
the full-depth DLS data (to push to a tighter shear threshold) and
getting spectroscopic redshifts for all the additional X-ray sources.

There are two trends evident in Table~\ref{table-xray}.  First, there
is a clear correlation between X-ray flux and lensing rank.  This is
not surprising given that we expect each of these attributes to be
linked to the more fundamental quantity, mass, but it is nevertheless
reassuring that this correlation is evident even before taking the
important next step of correcting for redshift effects, that is,
translating $F_X$ into $L_X$ and translating shear into mass.  The
relation between lensing mass and X-ray luminosity (and temperature)
will be extremely interesting, as it addresses the relation between
observable and predictable quantities for X-ray cluster surveys.  The
same is true of optical M/L in the context of optical cluster
selection.

Second, each of the three X-ray clusters judged to be unrelated to the
main candidate is the lowest-ranked X-ray source in its target area.
That is, the shear selection is not missing clusters which would be
judged to be massive on the basis of an X-ray survey.  Again, this
statement hides some complexity, as one must first infer a mass from
the X-ray data and then ask whether it should have been detected by
lensing given its redshift.  Nevertheless, this trend does suggest
that the two methods reveal mostly overlapping subsets of the cluster
population.  The case which comes closest to being an exception to
this trend would appear to be candidate 8 (DLSCL J1055.2-0503), where
the secondary source is only $\sim 25\%$ X-ray fainter than the
primary target and at a similar redshift.  But the primary candidate
is very near the threshold for lensing selection, so a $\sim 25\%$
less massive neighbor {\it should} be below the lensing threshold.

\subsection{Correspondence with Optical Galaxy Clusters}

The shear-selected clusters in this sample overwhelmingly correspond
to clusters of galaxies in the optical imaging; there are no
qualitatively ``dark'' mass concentrations.  This will be further
explored with quantitative measurements, such as optical mass-to-light
ratios, in a future paper.  Even in the case of DLSCL J1402.2-1028,
which is likely to be a projection based on the X-ray and
spectroscopic evidence, there appears to be a galaxy overdensity at
the (two-dimensional) position of the shear-selected cluster
candidate.  This is a consequence of lensing and
(two-dimensional) galaxy density having similar properties under
line-of-sight projection, in contrast with the other two lines of
evidence.

Where X-rays are detected, the position of the extended X-ray emission
always matches that of the galaxy overdensity, which in many cases is
different from the lensing peak position.  The most likely explanation
is uncertainty in the lensing position.  As a guide to the
uncertainty, the lensing position of DLSCL J1054.1-0549 changed by
$\sim$1\arcm\ when additional imaging data became available and a new
stack was made.  Also, simulations show that for data of this depth,
while the typical offset between the true and measured lens centers
due to statistical noise is 20-30\arcs, offsets of up to 100\arcs\ can
occur.  There is an additional digitization noise from the 30\arcs\
pixels in the convergence maps.  In short, offsets of up to
$\sim$2\arcm\ may not be significant in the current dataset.  This low
precision will be improved when the full-depth DLS data are analyzed.
In addition to the argument from the data, it is difficult to imagine
a scenario giving rise to real offsets between the dark matter and the
galaxies.  Even in a disruptive scenario such as a merger, both dark
matter and galaxies are collisionless, so they should remain together
while they separate from the collisional X-ray-emitting gas.

\subsection{Multiplicity}

Now that redshift information is available, we revisit the candidate
splitting criterion.  Because redshift information provides most of
the cosmology-constraining power of a cluster-counting survey, a clump
at a different redshift should be counted separately.  It need not
have been detectable in the absence of its projected neighbor, because
the comparison n-body mock survey will have cases of boosting by
projected neighbors as well.  Abell 781C, which probably would have
been detected in the absence of its neighbors anyway, should count as
a separate cluster.

Whether one splits clumps at the same redshift, such as Abell 781 A
and B, is a matter of taste.  Each would likely have been detected in
the absence of the other.  For purposes of cosmological constraints,
the precise definition is not important as long as the same criteria
are applied in the real survey and the mock survey from n-body
simulations.  The same is true for purposes of comparison to
optically or X-ray selected samples.

In short, adding redshift information yields an additional candidate
split from Abell 781 proper, for a total of nine candidates.

\subsection{Redshift Distribution}

Table~\ref{table-xray} includes the redshifts of the cluster
candidates.  They cover a broad range from 0.19 to 0.68 with a median
of $\sim0.35$.  This is the product of at least three factors: the
source redshift distribution, which determines the lensing efficiency
and thus the mass threshold as a function of lens redshift; the
cluster mass function, which determines how many clusters are above
the mass threshold at each redshift; and the volume probed as a
function of redshift.  For sources at $z=1$, the lensing efficiency
peaks for lenses at $z \sim 0.35$, and is a factor of two less for
lenses at $z \sim 0.1$ and $z \sim 0.7$.  At lens redshifts greater
than the efficiency peak, the higher mass threshold may be partially
compensated by the larger volume probed, but at low redshifts, the two
effects conspire to severely limit the expected number of lenses at $z
\sim 0.1$.  This is entirely consistent with the range seen here,
which (at the risk of over-interpreting small-number statistics) seems
to peak at $z \sim 0.3$ and have a long, non-negligible tail out to $z
\sim 0.7$, with a sharper cutoff toward low redshifts.

This is a qualitative picture, but details will matter when the sample
size grows large enough to fully characterize the observed lens
redshift distribution.  In particular, the sources cover a large
redshift range, rather than lying in a single plane, and this will
stretch the redshift range of the shear-selected clusters.  Accurate
characterization of the source redshift distribution, {\it given all
observational effects and cuts}, will be critical for doing precision
cosmology with large samples.  In addition, ``mass threshold'' is an
oversimplification of the true selection, because mass {\it profile}
plays a large role in determining detectability.  If mass profiles
change with redshift, this will feed through to the redshift
distribution.  This latter issue can be addressed by comparing to mock
lensing surveys of n-body simulations, so that no profile need be
assumed for purposes of comparison.

\subsection{Projections}

Hennawi \& Spergel (2004) have shown through simulations that
shear-selected samples will always have false positives, even for very
high shear thresholds.  These are caused by large-scale structure
noise, which cannot be overcome by improved observations.  The same
authors investigated the possibility of candidate followup with shear
tomography, which examines the growth of shear with source redshift.
This technique had been applied successfully by Wittman \etal\
(2001,2003) to constrain lens redshifts (to within $\sim 0.1$)
independent of any information about the lensing cluster.
Surprisingly, they found examples of false shear-selected candidates
which displayed perfectly sensible shear-redshift curves.  Although it
is not clear how often it fails, tomography is not infallible, and it
is natural to examine other followup possibilities to weed out the
false positives.

Unfortunately, other forms of followup, such as optical
imaging/spectroscopy of cluster member galaxies or X-ray/SZE detection
of the hot intracluster medium, bring back into the sample the very
biases which shear selection seeks to avoid.  If nature is kind, these
re-introduced biases will be small.  For example, even if
optically-selected samples are biased against low M/L systems, the low
M/L systems which are selectable by shear only may be confirmable with
spectroscopy.  A bias against extremely dark systems will remain to
some extent, but it will be much reduced.

If only one type of followup is pursued, spectroscopy is the natural
choice.  A cluster redshift, which neither X-ray nor SZE is likely to
provide, is always desirable, and photometric redshifts are not
accurate enough to conclusively rule out projections.  For the moment,
though, it is prudent to pursue as many forms of followup as possible,
to confirm the very tentative conclusions here.

If one is using clusters for cosmological constraints rather than as
astrophysical laboratories in their own right, there also remains the
possibility of simply counting shear peaks without attaching a
redshift to each one, or verifying that is a true overdensity in three
dimensions.  Cosmological constraints could be derived by comparison
with similar mock surveys of n-body simulation.  However, little work
has been done in this area theoretically because most of the
constraining power of cluster counts lies in the redshift
distribution.  Counting shear peaks without attached redshifts is a
way of examining the non-Gaussian properties of the cosmic shear
field, which in other contexts has been shown to supply cosmolgical
constraints independent of and complementary to those provided by the
Gaussian properties of the cosmic shear field (Jarvis \etal\ 2004).

\subsection{Extensions of this Work}

Many aspects of this investigation will be extended in forthcoming
papers: measurement of X-ray temperatures $T_X$ and the $T_X-L_X$
relation; shear calibration and relations between lensing mass and
$T_X$, $L_X$, and optical M/L; extension of the selection to the full
area covered by the DLS; investigation of the offsets between lensing
peaks and galaxy/X-ray peaks; production of a comparison
optically-selected sample from the DLS imaging data; lensing
tomography of the clusters, especially of DLSCL J1402.2-1028 to
determine if tomography indicates that it is a projection; and shear
selection with different filters, such as the tomographic matched
filter suggested by Hennawi \& Spergel (2004).

In the more distant future, large surveys such as LSST will find on
the order of 100,000 shear-selected clusters (Tyson \etal\ 2003).
With such massive statistics, systematics will become extremely
important.  Comparison to mock surveys of n-body simulations must
be done very carefully to avoid introducing systematics.  An efficient
way of dealing with projections must be in place, as spectroscopy and
X-ray followup are likely to be impractical on this scale.  The modest
samples produced by the DLS and other near-future surveys will provide
testbeds for developing these methods.

\acknowledgments 

We thank Matt Auger and Chris Fassnacht for assistance in reducing
LRIS spectroscopy, Perry Gee for providing the redshift of CXOU
J105535-045930, and Jim Bosch for assistance with preparing figures.
The DLS has received generous support from Lucent Technologies and
from NSF grants AST 04-41072 and AST 01-34753.  Support for this work
was also provided by the National Aeronautics and Space Administration
through {\it Chandra} Award Numbers GO3-4173A and GO3-4173B issued to
Rutgers and UC Davis by the {\it Chandra} X-ray Center, which is
operated by the Smithsonian Astrophysical Observatory for and on
behalf of the National Aeronautics Space Administration under contract
NAS8-03060.  We also thank NOAO for supporting survey programs, and
the 2dFGRS project for making data publicly available.  Observations
were obtained at Cerro Tololo Inter-American Observatory and the
W. M. Keck Observatory.  CTIO is a division of National Optical
Astronomy Observatory (NOAO), which is operated by the Association of
Universities for Research in Astronomy, Inc., under cooperative
agreement with the National Science Foundation.  This work also made
use of the Image Reduction and Analysis Facililty (IRAF), the
NASA/IPAC Extragalactic Database (NED), the NASA Astrophysics Data
System (ADS), and SAOImage DS9, developed by Smithsonian Astrophysical
Observatory.

\end{document}